\documentclass[%
 reprint,
 amsmath,amssymb,
 aps,
 pra,
]{revtex4-2}

\usepackage{graphicx}
\usepackage{dcolumn}
\usepackage{bm}

\usepackage{amsmath}
\usepackage{amssymb}
\usepackage{braket}
\usepackage{mathrsfs}
\usepackage{hyperref}
\usepackage{xfrac}
\usepackage{upgreek}

\renewcommand\bra[1]{{\langle{#1}|}}
\makeatletter
\renewcommand\ket[1]{%
  \@ifnextchar\bra{\k@t{#1}\!}{\k@t{#1}}%
}
\newcommand\k@t[1]{{|{#1}\rangle}}
\makeatother

\begin{document}

\preprint{APS/123-QED}

\title{Spontaneous optical emission of a randomly oriented chiral quantum source}
\author{Mikael P. Backlund}
 \email{mikaelb@illinois.edu}
\affiliation{%
 Department of Chemistry,
 Illinois Quantum Information Science and Technology Center (IQUIST), and Center for Biophysics and Quantitative Biology, University of Illinois at Urbana-Champaign, Urbana, IL, USA 61801
}%

\date{\today}

\begin{abstract}
We derive the form of the one-photon state of the electromagnetic field resulting from spontaneous optical emission of a randomly oriented chiral quantum source, including electric dipole, magnetic dipole, and electric quadrupole contributions to the radiation. We describe how spatial coherence in the emission can be exploited to reveal hidden information about the emitter. This work serves as a prelude and companion piece to our concurrently submitted work \cite{Backlund2026Letter} in which we present the classical and quantum error bounds associated with assigning the handedness of such an emitter.
\end{abstract}

\maketitle


\section{Introduction}
Scientists have been studying the preferential interactions between circularly polarized light and chiral molecules for well over a century \cite{barron2004molecular,berova2012comprehensive}. Harnessing chiroptical emission and absorption is a major goal in the development of numerous promising technologies \cite{vanorman2025chiral}. Biology is exquisitely sensitive to chirality \cite{blackmond2010origin}, and accurate enantiomeric identification and purification is profoundly important for applications in pharmaceuticals  \cite{tokunaga2018understanding,nguyen2006chiral}. Chiroptical spectroscopies comprise some of the most widely employed analytical techniques for distinguishing a molecule from its mirror-image counterpart, despite being fundamentally difficult due to the mismatch between optical wavelengths and molecular dimensions \cite{craig1998molecular}.

In a companion Letter we present fundamental speed limits to chiroptical discrimination from the perspective of quantum hypothesis testing \cite{Backlund2026Letter}. We confirm that simply counting $L$ vs. $R$ photons is the best-possible classical strategy, but also find that an optimal quantum measurement can in principle vastly improve the error probability given a fixed photon budget. Our starting point in that work is the one-photon state of the quantum electromagnetic field resulting from spontaneous emission of a randomly oriented quantum source. In the present Article, we will derive this density operator from first principles. The key feature is the mutual coherence between modes of different directions and polarizations, which can be understood from both a quantum and classical perspective. We discuss how coherent measurement of light collected from multiple directions around the emitter permits the disambiguation of molecular parameters that would otherwise be left unresolved by a conventional measurement of circular polarized luminescence (CPL) applied to an unoriented sample.

\section{Results and Discussion}
\subsection{Multipolar quantum emission}
\noindent We begin by writing the multipolar Hamiltonian describing the quantum field and molecule \cite{craig1998molecular}:
\begin{equation}
    H = H_\text{mol} + H_\text{rad} + H_\text{int},
\end{equation}
where $H_\text{mol}$ is the Hamiltonian of the bare molecule located at the origin, $H_\text{rad}$ is the Hamiltonian of the field, and $H_\text{int}$ describes their interaction. $H_\text{rad}$ is given in the usual modal decomposition by
\begin{equation} \label{eq_Hrad}
    H_\text{rad} = \sum_{\mathbf{k}s}\hbar\omega_k \hat{a}^\dagger(\mathbf{k}s)\hat{a}(\mathbf{k}s)
\end{equation}
where we omit the zero-point energy contribution to the field's energy. In Eq. (\ref{eq_Hrad}), $\omega_k=ck$ and $\hat{a}(\mathbf{k}s)$ is the annihilation operator for the mode corresponding to propagation vector $\mathbf{k}$ and polarization index $s$ satisfying the canonical commutation relation $[\hat{a}(\mathbf{k}s),\hat{a}^\dagger(\mathbf{k'}s')] = \delta_{\mathbf{k}\mathbf{k'}}\delta_{ss'}$. The multipolar interaction is truncated to include only contributions from the electric dipole, electric quadrupole, and magnetic dipole:
\begin{equation}
    H_\text{int} = \left\{-\frac{1}{\varepsilon_0}\mu_i d^\perp_i(\mathbf{r}) -\frac{1}{\varepsilon_0}Q_{ij}\nabla_j d^\perp_i(\mathbf{r}) - m_i b_i(\mathbf{r}) \right\}\Bigg|_{\mathbf{r}=\mathbf{0}},
\end{equation}
where $\boldsymbol{\upmu}$, $\mathbf{Q}$, and $\mathbf{m}$ are the electric dipole moment, electric quadrupole moment, and magnetic dipole moment operators for the molecule, respectively. We take the Einstein summation convention here and throughout. The microscopic displacement field operator is
\begin{eqnarray}
    \mathbf{d^\perp}(\mathbf{r}) = i \sum_{\mathbf{k}s}\sqrt{\frac{\hbar ck \varepsilon_0}{2V}}\Bigg\{\mathbf{e}\left(\mathbf{\hat{k}}s\right) \hat{a}(\mathbf{k}s)e^{i\mathbf{k}\cdot\mathbf{r}} \nonumber \\ - \mathbf{e}^*\left(\mathbf{\hat{k}}s\right) \hat{a}^\dagger(\mathbf{k}s)e^{-i\mathbf{k}\cdot\mathbf{r}} \Bigg\},
\end{eqnarray}
\begin{figure}
    \centering
    \includegraphics[width=\linewidth]{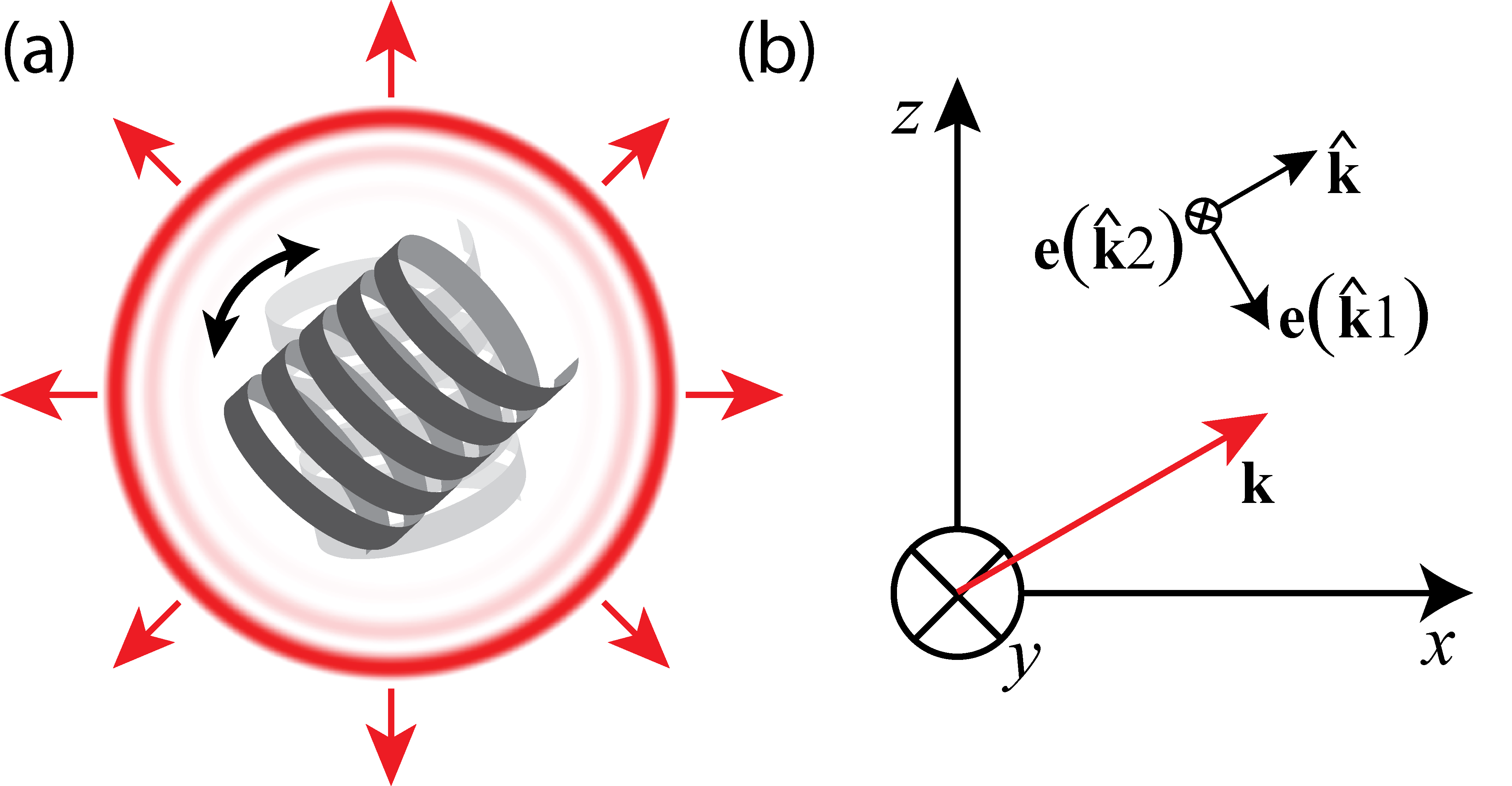}
    \caption{(a) Illustration of the model. We consider the spherically symmetric field due to a randomly oriented chiral quantum emitter, here depicted as a tumbling helix. (b) Lab-frame coordinate definitions, with wavevector $\mathbf{k}$. The unit vectors $\mathbf{e}\left(\hat{\mathbf{k}}1\right)$, $\mathbf{e}\left(\hat{\mathbf{k}}2\right)$, and $\hat{\mathbf{k}}$ form a right-handed triad.}
    \label{fig:overview}
\end{figure}
where $V$ is the quantization volume, $\mathbf{e}\left(\mathbf{\hat{k}}s\right)$ is the complex unit polarization vector of the corresponding mode, and $\hat{\mathbf{k}} = \mathbf{k}/k$. The magnetic field operator is
\begin{eqnarray}
    \mathbf{b}(\mathbf{r})=i\sum_{\mathbf{k}s}\sqrt{\frac{\hbar k}{2\varepsilon_0 c V}}\Bigg\{ \mathbf{b}\left(\mathbf{\hat{k}}s\right) \hat{a}(\mathbf{k}s)e^{i\mathbf{k}\cdot\mathbf{r}} \nonumber \\ - \mathbf{b}^*\left(\mathbf{\hat{k}}s\right) \hat{a}^\dagger(\mathbf{k}s)e^{-i\mathbf{k}\cdot\mathbf{r}} \Bigg\},
\end{eqnarray}
with $\mathbf{b}\left(\mathbf{\hat{k}}s\right) = \hat{\mathbf{k}} \times \mathbf{e}\left(\mathbf{\hat{k}}s\right)$. Going into the interaction picture with respect to $H_0 = H_\text{mol}+H_\text{rad}$ gives:
\begin{eqnarray}
    H'_\text{int}(t) = \Bigg\{-\frac{1}{\varepsilon_0}\mu_i(t) d^\perp_i(\mathbf{r},t) -\frac{1}{\varepsilon_0}Q_{ij}(t)\nabla_j d^\perp_i(\mathbf{r},t)\nonumber \\  - m_i(t) b_i(\mathbf{r},t) \Bigg\}\Bigg|_{\mathbf{r}=\mathbf{0}},
\end{eqnarray}
with
\begin{subequations}
    \begin{equation}
        \boldsymbol{\upmu}(t) = e^{iH_\text{mol}t/\hbar} \, \boldsymbol{\upmu} \, e^{-iH_\text{mol}t/\hbar}
    \end{equation}
    \begin{equation}
        \mathbf{Q}(t) = e^{iH_\text{mol}t/\hbar} \, \mathbf{Q} \, e^{-iH_\text{mol}t/\hbar}
    \end{equation}
    \begin{equation}
        \mathbf{m}(t) = e^{iH_\text{mol}t/\hbar} \, \mathbf{m} \, e^{-iH_\text{mol}t/\hbar}
    \end{equation}
\end{subequations}
and
\begin{subequations} \label{eq_fieldoperators}
    \begin{eqnarray}
        \mathbf{d^\perp}(\mathbf{r},t) = i \sum_{\mathbf{k}s}\sqrt{\frac{\hbar ck \varepsilon_0}{2V}}\Bigg\{ \mathbf{e}\left(\mathbf{\hat{k}}s\right) \hat{a}(\mathbf{k}s)e^{i\left(\mathbf{k}\cdot\mathbf{r}-\omega_kt\right)}\nonumber \\  - \mathbf{e}^*\left(\mathbf{\hat{k}}s\right) \hat{a}^\dagger(\mathbf{k}s)e^{-i\left(\mathbf{k}\cdot\mathbf{r}-\omega_kt\right)} \Bigg\}
    \end{eqnarray}
    \begin{eqnarray}
        \mathbf{b}(\mathbf{r},t)=i\sum_{\mathbf{k}s}\sqrt{\frac{\hbar k}{2\varepsilon_0 cV}}\Bigg\{ \mathbf{b}\left(\mathbf{\hat{k}}s\right) \hat{a}(\mathbf{k}s)e^{i\left(\mathbf{k}\cdot\mathbf{r}-\omega_kt\right)} \nonumber \\- \mathbf{b}^*\left(\mathbf{\hat{k}}s\right) \hat{a}^\dagger(\mathbf{k}s)e^{-i\left(\mathbf{k}\cdot\mathbf{r}-\omega_kt\right)} \Bigg\}.
    \end{eqnarray}
\end{subequations}

We take the initial state of the total system to be $\ket{1}\ket{\text{vac}}$, where $\ket{1}$ is the electronic excited state of the molecule and $\ket{\text{vac}}$ denotes the electromagnetic vacuum. Left to evolve under $H'_\text{int}(t)$, the molecule will eventually relax to its ground electronic state $\ket{0}$, leaving the field in a one-photon state. We are interested in the form of this one-photon state. We set up the problem in a manner akin to the well-established treatment due originally to Weisskopf and Wigner for the spontaneous emission of a two-level atom \cite{Weisskopf1930,scully1997quantum}. In treating our molecule as a two-level emitter we effectively neglect vibronic couplings \cite{schatz2002quantum} and other sources of molecular decoherence. Such a strong simplification is rationalized by the fact that chiroptical spectroscopies are typically performed in a wavelength-resolved manner \cite{barron2004molecular}, and that narrow spectral filtering can restore photon indistinguishability even from remote, incoherent sources \cite{deng2019quantum}.

At time $t$ we assume the state of the system to be a superposition of the form
\begin{equation} \label{eq_Psioft}
    \ket{\Psi(t)} = c_1(t)\ket{1}\ket{\text{vac}} + \sum_{\mathbf{k}s} c_{\mathbf{k}s}(t) \ket{0}\ket{1(\mathbf{k}s)},
\end{equation}
with $\ket{1(\mathbf{k}s)} = \hat{a}^\dagger(\mathbf{k}s)\ket{\text{vac}}$, $c_1(0)=1$, and $c_{\mathbf{k}s}(0)=0 \quad \forall \mathbf{k},s$. The relevant Schr\"{o}dinger equation is 
\begin{equation}
    \ket{\dot{\Psi}(t)} = -\frac{i}{\hbar} H'_\text{int}(t) \ket{\Psi(t)},
\end{equation}
which yields the coupled differential equations:
\begin{subequations} \label{eq_ODEs}
    \begin{equation}
        \dot{c}_1(t) = -\frac{i}{\hbar} \sum_{\mathbf{k}s} c_{\mathbf{k}s}(t)\bra{\text{vac}}\bra{1}H'_\text{int}(t)\ket{0}\ket{1(\mathbf{k}s)}
    \end{equation}
    \begin{equation}
        \dot{c}_{\mathbf{k}s}(t) = -\frac{i}{\hbar} c_1(t) \bra{1(\mathbf{k}s)}\bra{0} H'_\text{int}(t)\ket{1}\ket{\text{vac}}.
    \end{equation}
\end{subequations}
The relevant matrix element is:
\begin{widetext}
\begin{equation}
    \bra{1(\mathbf{k}s)}\bra{0} H'_\text{int}(t)\ket{1}\ket{\text{vac}} = i\sqrt{\frac{\hbar c k}{2 \varepsilon_0 V}} e^{i(\omega_k-\omega_0)t} \left( \mu_i^{01}e^*_i\left(\mathbf{\hat{k}}s\right) - i k_j Q_{ij}^{01} e^*_i\left(\mathbf{\hat{k}}s\right) + \frac{1}{c} m_i^{01}b^*_i\left(\mathbf{\hat{k}}s\right)\right),
\end{equation}
\end{widetext}
where $\hbar\omega_0$ is the energy splitting between the molecular states and
\begin{subequations}
    \begin{equation}
        \boldsymbol{\upmu^{01}} = \braket{0|\boldsymbol{\upmu}|1}
    \end{equation}
    \begin{equation}
        \mathbf{Q^{01}} = \braket{0|\mathbf{Q}|1}
    \end{equation}
    \begin{equation}
        \mathbf{m^{01}} = \braket{0|\mathbf{m}|1}.
    \end{equation}
\end{subequations}
Defining:
\begin{eqnarray} \label{eq_gdef}
    g(\mathbf{k}s) \equiv&& i\sqrt{\frac{c k}{2 \hbar \varepsilon_0 V}} \Bigg( \mu_i^{01}e^*_i\left(\mathbf{\hat{k}}s\right) \nonumber \\ &&- i k_j Q_{ij}^{01} e^*_i\left(\mathbf{\hat{k}}s\right) + \frac{1}{c} m_i^{01}b^*_i\left(\mathbf{\hat{k}}s\right)\Bigg),
\end{eqnarray}
Eq. (\ref{eq_ODEs}) becomes:
\begin{subequations} \label{eq_ODEs2}
    \begin{equation}
        \dot{c}_1(t) = -i \sum_{\mathbf{k}s} c_{\mathbf{k}s}(t) \, g^*(\mathbf{k}s) e^{-i(\omega_k-\omega_0)t}
    \end{equation}
    \begin{equation}
        \dot{c}_{\mathbf{k}s}(t) = -i c_1(t) \, g(\mathbf{k}s) e^{i(\omega_k-\omega_0)t}.
    \end{equation}
\end{subequations}
Integrating Eq. (\ref{eq_ODEs2}b) gives:
\begin{equation} \label{eq_mycoeff}
    c_{\mathbf{k}s}(t) = -i g(\mathbf{k}s) \int_0^t \mathrm{d}t' e^{i(\omega_k-\omega_0)t'} c_1(t'),
\end{equation}
which can be substituted into Eq. (\ref{eq_ODEs2}a) to yield:
\begin{equation} \label{eq_c1dotintdiff}
    \dot{c}_1(t) = - \sum_{\mathbf{k}s} \left|g(\mathbf{k}s)\right|^2 \int_0^t \mathrm{d}t' e^{i(\omega_k-\omega_0)(t'-t)} c_1(t').
\end{equation}
One could solve for $c_1(t)$ and in turn for the spectrum of the emitted photon at various levels of approximation \cite{Weisskopf1930,scully1997quantum,cohentannoudji1992atomphoton,Berman_PRA_2010}-- the only wrinkle here is that we include the magnetic dipole and electric quadrupole contributions to the emission in addition to the usual electric dipole. Since in this work we are primarily interested in the geometrical rather than spectral distribution of the emission, however, we proceed slightly differently.

Substituting Eq. (\ref{eq_mycoeff}) into Eq. (\ref{eq_Psioft}) gives:
\begin{eqnarray}
    \ket{\Psi(t)} =&& c_1(t)\ket{1}\ket{\text{vac}} \nonumber \\ &&- i \sum_{\mathbf{k}s} g(\mathbf{k}s) \int_0^t \mathrm{d}t' e^{i(\omega_k-\omega_0)t'} c_1(t') \ket{0}\ket{1(\mathbf{k}s)} \nonumber \\ =&& c_1(t)\ket{1}\ket{\text{vac}} \nonumber \\ &&+ \sum_{\mathbf{k}s} g(\mathbf{k}s) \tilde{\Phi}(k,t)\ket{0}\ket{1(\mathbf{k}s)},
\end{eqnarray}
where we've absorbed factors such that $\tilde{\Phi}(k,t)$ depends only on the wavenumber of the mode and not on its direction or polarization. Tracing out the molecular degrees of freedom leaves the field in the state:
\begin{widetext}
\begin{equation}
    \rho(t) = c_1(t)c_1^*(t)\ket{\text{vac}}\bra{\text{vac}} + \sum_{\mathbf{k}s}\sum_{\mathbf{k'}s'} \tilde{\Phi}(k,t) \tilde{\Phi}^*(k',t) g(\mathbf{k}s)g^*(\mathbf{k'}s') \ket{1(\mathbf{k}s)}\bra{1(\mathbf{k'}s')}.
\end{equation}
\end{widetext}
In measurements such as the one we consider here, the number of photons emitted (or detected) is typically the important resource. We thus post-select (e.g. by waiting long enough) for the one-photon state of the field:
\begin{equation}
    \rho = \mathcal{N} \sum_{\mathbf{k}s}\sum_{\mathbf{k'}s'} \Phi(k) \Phi^*(k') g(\mathbf{k}s)g^*(\mathbf{k'}s') \ket{1(\mathbf{k}s)}\bra{1(\mathbf{k'}s')},
\end{equation}
where we've defined $\Phi(k) \equiv \tilde{\Phi}(k,t\to\infty)$ and introduced the normalization constant $\mathcal{N}$. At this point we could phenomenologically incorporate decoherence arising from dephasing of the molecule by suitable averaging of $\Phi(k)\Phi^*(k')$ \cite{MitchellPRA2023}, but we choose not to do so here for the reasons explained above. In the event that the molecular orientation is uniformly random it is appropriate to take the rotational average denoted by angular brackets:
\begin{equation}
    \bar{\rho} = \mathcal{N}\sum_{\mathbf{k}s}\sum_{\mathbf{k'}s'} \Phi(k)\Phi^*(k') \langle g(\mathbf{k}s)g^*(\mathbf{k'}s') \rangle \, \ket{1(\mathbf{k}s)}\bra{1(\mathbf{k'}s')}.
\end{equation}
Next we make use of the tensorial relations in Appendix \ref{appendix_rotationaavs} to compute this rotational average. Referring to Eq. (\ref{eq_gdef}) and expanding we have:
\begin{widetext}
\begin{eqnarray} \label{eq_rotavg1}
    \langle g(\mathbf{k}s) g^*(\mathbf{k'}s')\rangle = \frac{c \sqrt{kk'}}{2 \hbar \varepsilon_0 V} \Bigg( \langle \mu_i^{01} \mu_{i'}^{01}\rangle e^*_i\left(\mathbf{\hat{k}}s\right)e_{i'}\left(\mathbf{\hat{k}'}s'\right) + ik'_{j'}\langle \mu_i^{01} Q_{i'j'}^{01}\rangle e^*_i\left(\mathbf{\hat{k}}s\right)e_{i'}\left(\mathbf{\hat{k}'}s'\right) \nonumber \\ - \frac{1}{c}\langle \mu_i^{01} m_{i'}^{01} \rangle e^*_i\left(\mathbf{\hat{k}}s\right)b_{i'}\left(\mathbf{\hat{k}'}s'\right) \nonumber \\ -i k_j\langle Q_{ij}^{01} \mu_{i'}^{01}\rangle e^*_i\left(\mathbf{\hat{k}}s\right)e_{i'}\left(\mathbf{\hat{k}'}s'\right) + k_jk'_{j'}\langle Q_{ij}^{01} Q_{i'j'}^{01}\rangle e^*_i\left(\mathbf{\hat{k}}s\right)e_{i'}\left(\mathbf{\hat{k}'}s'\right) \nonumber \\ + \frac{i}{c} k_j\langle Q_{ij}^{01} m_{i'}^{01} \rangle e^*_i\left(\mathbf{\hat{k}}s\right)b_{i'}\left(\mathbf{\hat{k}'}s'\right) \nonumber \\ + \frac{1}{c} \langle m_i^{01} \mu_{i'}^{01}\rangle b^*_i\left(\mathbf{\hat{k}}s\right)e_{i'}\left(\mathbf{\hat{k}'}s'\right) + \frac{i}{c}k'_{j'}\langle m_i^{01} Q_{i'j'}^{01}\rangle b^*_i\left(\mathbf{\hat{k}}s\right)e_{i'}\left(\mathbf{\hat{k}'}s'\right) \nonumber \\ - \frac{1}{c^2} \langle m_i^{01} m_{i'}^{01} \rangle b^*_i\left(\mathbf{\hat{k}}s\right)b_{i'}\left(\mathbf{\hat{k}'}s'\right) \Bigg),
\end{eqnarray}
\end{widetext}
where we've assumed that the elements of $\boldsymbol{\upmu^{01}}$ and $\mathbf{Q^{01}}$ are real and those of $\mathbf{m^{01}}$ are purely imaginary \cite{wakabayashi2014anisotropic}. The elements of these tensors can be related to those defined with respect to a molecule-fixed coordinate system via:
\begin{subequations}
    \begin{equation}
        \mu_i^{01} = R_{i\lambda} \mu_\lambda^{01}
    \end{equation}
    \begin{equation}
        m_i^{01} = R_{i\lambda} m_\lambda^{01}
    \end{equation}
    \begin{equation}
        Q_{ij}^{01} = R_{i\lambda} R_{j \mu} Q_{\lambda \mu}^{01},
    \end{equation}
\end{subequations}
where Greek indices indicate molecule-fixed coordinates, Latin indices indicate lab-fixed coordinates, and $\mathbf{R}$ is the matrix of direction cosines connecting the two coordinate systems. The various tensor averages are:
\begin{subequations}
    \begin{eqnarray}
        \langle \mu_i^{01}\mu_{i'}^{01} \rangle &=& \langle R_{i\lambda}R_{i'\lambda'} \rangle \mu_\lambda^{01} \mu_{\lambda'}^{01} \nonumber \\ &=& \frac{1}{3} \delta_{ii'}\delta_{\lambda\lambda'} \mu_\lambda^{01} \mu_{\lambda'}^{01} = \frac{1}{3}\delta_{ii'} \left| \mu^{01} \right|^2
    \end{eqnarray}
    \begin{equation}
        \langle m_i^{01}m_{i'}^{01} \rangle = -\frac{1}{3}\delta_{ii'} \left| m^{01} \right|^2
    \end{equation}
    \begin{equation}
        \langle \mu_i^{01}m_{i'}^{01} \rangle = \frac{i}{3}\delta_{ii'} \text{Im}\left(\boldsymbol{\upmu^{01}}\cdot\mathbf{m^{01}}\right)
    \end{equation}
    \begin{eqnarray}
        \langle \mu_i^{01} Q_{i'j'}^{01} \rangle &=& \langle R_{i\lambda} R_{i'\lambda'} R_{j'\mu'} \rangle \mu_\lambda^{01}Q_{\lambda'\mu'}^{01} \nonumber \\ &=& \frac{1}{6}\varepsilon_{ii'j'}\varepsilon_{\lambda\lambda'\mu'}\mu_\lambda^{01}Q_{\lambda'\mu'}^{01} = 0
    \end{eqnarray}
    \begin{eqnarray}
        \langle m_i^{01} Q_{i'j'}^{01} \rangle &=& \langle R_{i\lambda} R_{i'\lambda'} R_{j'\mu'} \rangle m_\lambda^{01}Q_{\lambda'\mu'}^{01} \nonumber \\ &=& \frac{1}{6}\varepsilon_{ii'j'}\varepsilon_{\lambda\lambda'\mu'}m_\lambda^{01}Q_{\lambda'\mu'}^{01} = 0
    \end{eqnarray}
    \begin{eqnarray}
        \langle Q_{ij}^{01}Q_{i'j'}^{01} \rangle&&= \langle R_{i\lambda}R_{j\mu}R_{i'\lambda'}R_{j'\mu'} \rangle Q_{\lambda \mu}^{01}Q_{\lambda'\mu'}^{01} \nonumber \\ &&= \frac{\left(\text{Tr}\,\mathbf{Q^{01}}\right)^2}{30}\left( 4\delta_{ij}\delta_{i'j'} - \delta_{ii'}\delta_{jj'}-\delta_{ij'}\delta_{i'j} \right) \nonumber \\&& + \frac{\left\lVert \mathbf{Q^{01}} \right\rVert^2_F}{30} \left( -2\delta_{ij}\delta_{i'j'} +3 \delta_{ii'}\delta_{jj'}+3\delta_{ij'}\delta_{i'j} \right), \nonumber \\
    \end{eqnarray}
\end{subequations}
where we have made use of the fact that $\mathbf{Q}^{01}$ is symmetric and the Frobenius norm of a matrix $\mathbf{A}$ is defined by $\left\lVert\mathbf{A}\right\rVert_F^2 = \sum_{ij}\left|A_{ij}\right|^2$. Equation (\ref{eq_rotavg1}) simplifies:
\begin{widetext}
\begin{eqnarray} \label{eq_rotavg2}
    \langle g(\mathbf{k}s)g^*(\mathbf{k'}s')\rangle &&= \frac{c \sqrt{kk'}}{2 \hbar \varepsilon_0 V} \Bigg( \frac{1}{3}\left|\mu^{01}\right|^2 e^*_i\left(\mathbf{\hat{k}}s\right)e_i\left(\mathbf{\hat{k}'}s'\right) + \frac{1}{3c^2} \left|m^{01}\right|^2 b^*_i\left(\mathbf{\hat{k}}s\right)b_i\left(\mathbf{\hat{k}'}s'\right) \nonumber \\ &&- \frac{i}{3c} \text{Im}\left(\boldsymbol{\upmu^{01}}\cdot\mathbf{m^{01}}\right) \left[ e^*_i\left(\mathbf{\hat{k}}s\right)b_i\left(\mathbf{\hat{k}'}s'\right) - b^*_i\left(\mathbf{\hat{k}}s\right)e_i\left(\mathbf{\hat{k}'}s'\right) \right] \nonumber \\ &&+ \frac{1}{30}\left(\text{Tr}\,\mathbf{Q^{01}}\right)^2 \left[ 4k_i k'_j e^*_i\left(\mathbf{\hat{k}}s\right)e_j\left(\mathbf{\hat{k}'}s'\right) - k_jk'_j e^*_i\left(\mathbf{\hat{k}}s\right)e_i\left(\mathbf{\hat{k}'}s'\right) - k'_i k_j e^*_i\left(\mathbf{\hat{k}}s\right) e_j\left(\mathbf{\hat{k}'}s'\right) \right] \nonumber \\ &&+ \frac{1}{30} \left\lVert\mathbf{Q^{01}}\right\rVert_F^2 \left[ -2k_i k'_j e^*_i\left(\mathbf{\hat{k}}s\right)e_j\left(\mathbf{\hat{k}'}s'\right) +3 k_jk'_j e^*_i\left(\mathbf{\hat{k}}s\right)e_i\left(\mathbf{\hat{k}'}s'\right) +3 k'_i k_j e^*_i\left(\mathbf{\hat{k}}s\right) e_j\left(\mathbf{\hat{k}'}s'\right) \right] \Bigg), \nonumber \\
\end{eqnarray}
\end{widetext}
and further to:
\begin{widetext}
\begin{eqnarray} \label{eq_rotavg3}
    \langle g(\mathbf{k}s) g^*(\mathbf{k'}s')\rangle &&= \frac{c \sqrt{kk'}}{2 \hbar \varepsilon_0 V} \Bigg( \left[\frac{1}{3}\left|\mu^{01}\right|^2 + \frac{k_jk'_j}{30}\left( 3\left\lVert\mathbf{Q^{01}}\right\rVert_F^2 - \left(\text{Tr}\,\mathbf{Q^{01}}\right)^2\right) \right] e^*_i\left(\mathbf{\hat{k}}s\right)e_i\left(\mathbf{\hat{k}'}s'\right) \nonumber \\ &&+ \frac{1}{3c^2} \left|m^{01}\right|^2 b^*_i\left(\mathbf{\hat{k}}s\right)b_i\left(\mathbf{\hat{k}'}s'\right) \nonumber \\ &&- \frac{i}{3c} \text{Im}\left(\boldsymbol{\upmu^{01}}\cdot\mathbf{m^{01}}\right) \left[ e^*_i\left(\mathbf{\hat{k}}s\right)b_i\left(\mathbf{\hat{k}'}s'\right) - b^*_i\left(\mathbf{\hat{k}}s\right)e_i\left(\mathbf{\hat{k}'}s'\right) \right] \nonumber \\ && + \left(3\left\lVert\mathbf{Q^{01}}\right\rVert_F^2 - \left(\text{Tr}\,\mathbf{Q^{01}}\right)^2 \right)\frac{k_jk'_i}{30} e^*_i\left(\mathbf{\hat{k}}s\right)e_j\left(\mathbf{\hat{k}'}s'\right) \Bigg),
\end{eqnarray}
\end{widetext}
where we've used the fact that $\mathbf{k}$ and $\mathbf{e}\left(\mathbf{\hat{k}}s\right)$ are mutually orthogonal. In subsequent expressions we consolidate our notation by defining:
\begin{equation}
    \mathcal{Q}\left(\mathbf{Q}^{01}\right) \equiv \frac{1}{10}\left(3\left\lVert\mathbf{Q}^{01}\right\rVert_F^2 -\left(\text{Tr}\,\mathbf{Q^{01}}\right)^2\right),
\end{equation}
as well as the optical rotary strength:
\begin{equation}
    \mathscr{R}^{01} \equiv \text{Im}\left(\boldsymbol{\upmu^{10}}\cdot\mathbf{m^{01}}\right),
\end{equation}
which is equal to 0 for an achiral molecule \cite{craig1998molecular}.
\subsection{Chiroptical quantum emission}
In this Article and the accompanying Letter \cite{Backlund2026Letter} we concern ourselves with the ability to discriminate the field states $\bar{\rho}_+$ and $\bar{\rho}_-$ resulting from emission due to stereoisomer pairs with:
\begin{eqnarray}
    \bar{\rho}_\pm =&& \mathcal{N}\sum_{\mathbf{k}s}\sum_{\mathbf{k'}s'} \Phi(k) \Phi^*(k')\times \nonumber \\ && \langle g_\pm(\mathbf{k}s)g_\pm^*(\mathbf{k'}s') \rangle \,\ket{1(\mathbf{k}s)}\bra{1(\mathbf{k'}s')}
\end{eqnarray}
and 
\begin{widetext}
\begin{eqnarray}
    \langle g_\pm(\mathbf{k}s) g_\pm^*(\mathbf{k'}s')\rangle &&= \frac{c \sqrt{kk'}}{2 \hbar \varepsilon_0 V} \Bigg( \left[\frac{1}{3}\left|\mu^{01}\right|^2 + \frac{1}{3}k_jk'_j \mathcal{Q}\left(\mathbf{Q^{01}}\right) \right] e^*_i\left(\mathbf{\hat{k}}s\right)e_i\left(\mathbf{\hat{k}'}s'\right) \nonumber \\ &&+ \frac{1}{3c^2} \left|m^{01}\right|^2 b^*_i\left(\mathbf{\hat{k}}s\right)b_i\left(\mathbf{\hat{k}'}s'\right) \nonumber \\ && \mp \frac{i}{3c} \mathscr{R}^{01} \left[ e^*_i\left(\mathbf{\hat{k}}s\right)b_i\left(\mathbf{\hat{k}'}s'\right) - b^*_i\left(\mathbf{\hat{k}}s\right)e_i\left(\mathbf{\hat{k}'}s'\right) \right] \nonumber \\ && + \frac{1}{3}\mathcal{Q}\left(\mathbf{Q^{01}}\right) k_jk'_i e^*_i\left(\mathbf{\hat{k}}s\right)e_j\left(\mathbf{\hat{k}'}s'\right) \Bigg).
\end{eqnarray}
\end{widetext}
To this point we have left the choice of polarization basis unspecified. Let's now designate the left/right circularly polarized basis $s\in\{L,R\}$. On the way to writing down the components of the circular polarization unit vectors we define the following linear polarization unit vectors (Fig. \ref{fig:overview}b) \cite{mandel1995optical}:
\begin{subequations}
    \begin{equation}
        \mathbf{e}\left(\hat{\mathbf{k}}1\right) = [\cos\theta_\mathbf{k}\cos\phi_\mathbf{k},\cos\theta_\mathbf{k}\sin\phi_\mathbf{k},-\sin\theta_\mathbf{k}]^\text{T}
    \end{equation}
    \begin{equation}
        \mathbf{e}\left(\hat{\mathbf{k}}2\right) = [-\sin\phi_\mathbf{k},\cos\phi_\mathbf{k},0]^\text{T},
    \end{equation}
\end{subequations}
where $\hat{\mathbf{k}} = [\cos\phi_\mathbf{k} \sin\theta_\mathbf{k},\sin\phi_\mathbf{k} \sin\theta_\mathbf{k},\cos\theta_\mathbf{k}]^\text{T}$ and $\mathbf{e}\left(\hat{\mathbf{k}}1\right), \mathbf{e}\left(\hat{\mathbf{k}}2\right), \hat{\mathbf{k}}$ form a right-handed triad. In terms of this linear polarization basis, the circular polarization basis vectors are:
\begin{subequations}
    \begin{equation}
        \mathbf{e}\left(\hat{\mathbf{k}}L\right) = \frac{1}{\sqrt{2}}\left[ \mathbf{e}\left(\hat{\mathbf{k}}1\right) + i\mathbf{e}\left(\hat{\mathbf{k}}2\right) \right]
    \end{equation}
    \begin{equation}
        \mathbf{e}\left(\hat{\mathbf{k}}R\right) = \frac{1}{\sqrt{2}}\left[ \mathbf{e}\left(\hat{\mathbf{k}}1\right) - i\mathbf{e}\left(\hat{\mathbf{k}}2\right) \right],
    \end{equation}
\end{subequations}
where we have taken the optics convention as in Ref. \cite{craig1998molecular}, for which the relations
\begin{subequations}\label{eq_ebrelationcirc}
    \begin{equation}
        b_i\left(\mathbf{\hat{k}}L\right) = -ie_i\left(\mathbf{\hat{k}}L\right)
    \end{equation}
    \begin{equation}
        b_i\left(\mathbf{\hat{k}}R\right) = ie_i\left(\mathbf{\hat{k}}R\right)
    \end{equation}
\end{subequations}
hold. Making use of Eq. (\ref{eq_ebrelationcirc}), the states $\bar{\rho}_\pm$ can be written:
\begin{widetext}
\begin{eqnarray}
    \bar{\rho}_\pm = \mathcal{N}\sum_{\mathbf{k},\mathbf{k'}} \Phi(k)\Phi^*(k')\frac{c\sqrt{kk'}}{6\hbar\varepsilon_0V} &&\Bigg\{\tilde{G}^{(\pm)}_{LL}(\mathbf{k},\mathbf{k'}) \ket{1(\mathbf{k}L)}\bra{1(\mathbf{k'}L)} + \tilde{G}^{(\pm)}_{RR}(\mathbf{k},\mathbf{k'}) \ket{1(\mathbf{k}R)}\bra{1(\mathbf{k'}R)} \nonumber \\ &&+ \tilde{G}_{LR}(\mathbf{k},\mathbf{k'})\ket{1(\mathbf{k}L)}\bra{1(\mathbf{k'}R)} + \tilde{G}_{RL}(\mathbf{k},\mathbf{k'})\ket{1(\mathbf{k}R)}\bra{1(\mathbf{k'}L)} \Bigg\}
\end{eqnarray}
where
\begin{subequations} \label{eq_Gtildedefs}
    \begin{eqnarray}
        \tilde{G}_{LL}^{(\pm)}(\mathbf{k},\mathbf{k'}) = \left[ \left(\mu^{01}\right)^2  \mp \frac{2}{c}\mathscr{R}^{01} + \frac{1}{c^2}\left|m^{01}\right|^2 + k k' \mathcal{Q}\left(\mathbf{Q^{01}}\right)\hat{k}_j\hat{k}'_j\right]e_i^*\left(\mathbf{\hat{k}}L\right)e_i\left(\mathbf{\hat{k}'}L\right) \nonumber \\ + kk'\mathcal{Q}\left(\mathbf{Q^{01}}\right) \hat{k}_j \hat{k}'_i e_i^*\left(\mathbf{\hat{k}}L\right)e_j\left(\mathbf{\hat{k}'}L\right)
    \end{eqnarray}
    \begin{eqnarray}
        \tilde{G}_{RR}^{(\pm)}(\mathbf{k},\mathbf{k'}) = \left[ \left(\mu^{01}\right)^2 \pm \frac{2}{c}\mathscr{R}^{01} + \frac{1}{c^2}\left|m^{01}\right|^2 + kk' \mathcal{Q}\left(\mathbf{Q^{01}}\right) \hat{k}_j\hat{k}'_j\right]e_i^*\left(\mathbf{\hat{k}}R\right)e_i\left(\mathbf{\hat{k}'}R\right) \nonumber \\ + kk'\mathcal{Q}\left(\mathbf{Q^{01}}\right) \hat{k}_j \hat{k}'_ie_i^*\left(\mathbf{\hat{k}}R\right)e_j\left(\mathbf{\hat{k}'}R\right)
    \end{eqnarray}
    \begin{eqnarray}
        \tilde{G}_{LR}(\mathbf{k},\mathbf{k'}) = \left[ \left(\mu^{01}\right)^2 - \frac{1}{c^2}\left|m^{01}\right|^2 + kk' \mathcal{Q}\left(\mathbf{Q^{01}}\right) \hat{k}_j\hat{k}'_j\right]e_i^*\left(\mathbf{\hat{k}}L\right)e_i\left(\mathbf{\hat{k}'}R\right) \nonumber \\ + kk'\mathcal{Q}\left(\mathbf{Q^{01}}\right) \hat{k}_j \hat{k}'_ie_i^*\left(\mathbf{\hat{k}}L\right)e_j\left(\mathbf{\hat{k}'}R\right)
    \end{eqnarray}
    \begin{eqnarray}
        \tilde{G}_{RL}(\mathbf{k},\mathbf{k'}) = \left[ \left(\mu^{01}\right)^2 - \frac{1}{c^2}\left|m^{01}\right|^2 + kk' \mathcal{Q}\left(\mathbf{Q^{01}}\right) \hat{k}_j\hat{k}'_j\right]e_i^*\left(\mathbf{\hat{k}}R\right)e_i\left(\mathbf{\hat{k}'}L\right) \nonumber \\ + kk'\mathcal{Q}\left(\mathbf{Q^{01}}\right) \hat{k}_j \hat{k}'_ie_i^*\left(\mathbf{\hat{k}}R\right)e_j\left(\mathbf{\hat{k}'}L\right).
    \end{eqnarray}
\end{subequations}
\end{widetext}
The normalization factor $\mathcal{N}$ can be specified at this point by setting $\text{Tr}\bar{\rho}_\pm=1$ and rearranging:
\begin{equation}
    \mathcal{N} = \frac{3\hbar \varepsilon_0 V}{c D^{01}\left(\sum_{\mathbf{k}}k\left|\Phi(k)\right|^2\right)},
\end{equation}
where we've introduced the generalized ``dipole'' strength \cite{gendron2019ab}, $D^{01}$, defined by:
\begin{equation} \label{eq_generalizeddipolestrength}
    D^{01} \equiv \left(\mu^{01}\right)^2 + \frac{1}{c^2}\left|m^{01}\right|^2 + \mathcal{Q}\left(\mathbf{Q^{01}}\right)\frac{\sum_\mathbf{k}k^3\left|\Phi(k)\right|^2}{\sum_\mathbf{k}k\left|\Phi(k)\right|^2}.
\end{equation}
These expressions allow us to rewrite:
\begin{widetext}
\begin{eqnarray}
    \bar{\rho}_\pm = \frac{1}{2 D^{01} \left(\sum_\mathbf{k}k\left|\Phi(k)\right|^2\right)}\sum_{\mathbf{k},\mathbf{k'}} \Phi(k)\Phi^*(k')\sqrt{k k'} &&\Bigg\{\tilde{G}^{(\pm)}_{LL}(\mathbf{k},\mathbf{k'}) \ket{1(\mathbf{k}L)}\bra{1(\mathbf{k'}L)} + \tilde{G}^{(\pm)}_{RR}(\mathbf{k},\mathbf{k'}) \ket{1(\mathbf{k}R)}\bra{1(\mathbf{k'}R)} \nonumber \\ &&+ \tilde{G}_{LR}(\mathbf{k},\mathbf{k'})\ket{1(\mathbf{k}L)}\bra{1(\mathbf{k'}R)} + \tilde{G}_{RL}(\mathbf{k},\mathbf{k'})\ket{1(\mathbf{k}R)}\bra{1(\mathbf{k'}L)} \Bigg\},
\end{eqnarray}
or, equivalently:
\begin{eqnarray} \label{eq_rhopmusefuldefn}
    \bar{\rho}_\pm = \frac{1}{2 \left(\sum_\mathbf{k}k\left|\Phi(k)\right|^2\right)}\sum_{\mathbf{k},\mathbf{k'}} \Phi(k)\Phi^*(k')\sqrt{k k'} &&\Bigg\{G^{(\pm)}_{LL}(\mathbf{k},\mathbf{k'}) \ket{1(\mathbf{k}L)}\bra{1(\mathbf{k'}L)} + G^{(\pm)}_{RR}(\mathbf{k},\mathbf{k'}) \ket{1(\mathbf{k}R)}\bra{1(\mathbf{k'}R)} \nonumber \\ &&+ G_{LR}(\mathbf{k},\mathbf{k'})\ket{1(\mathbf{k}L)}\bra{1(\mathbf{k'}R)} + G_{RL}(\mathbf{k},\mathbf{k'})\ket{1(\mathbf{k}R)}\bra{1(\mathbf{k'}L)} \Bigg\},
\end{eqnarray}
where:
\begin{subequations} \label{eq_Gdefs}
    \begin{equation}
        G_{LL}^{(\pm)}(\mathbf{k},\mathbf{k'}) = \left[ \epsilon_\mu  \mp \frac{\mathscr{G}}{2} + \epsilon_m + \tilde{\epsilon}_Qk k'\hat{k}_j\hat{k}'_j\right]e_i^*\left(\mathbf{\hat{k}}L\right)e_i\left(\mathbf{\hat{k}'}L\right) + \tilde{\epsilon}_Q kk'\hat{k}_j \hat{k}'_i e_i^*\left(\mathbf{\hat{k}}L\right)e_j\left(\mathbf{\hat{k}'}L\right)
    \end{equation}
    \begin{equation}
        G_{RR}^{(\pm)}(\mathbf{k},\mathbf{k'}) = \left[ \epsilon_\mu \pm \frac{\mathscr{G}}{2} + \epsilon_m + \tilde{\epsilon}_Q kk' \hat{k}_j\hat{k}'_j\right]e_i^*\left(\mathbf{\hat{k}}R\right)e_i\left(\mathbf{\hat{k}'}R\right) + \tilde{\epsilon}_Q kk' \hat{k}_j \hat{k}'_ie_i^*\left(\mathbf{\hat{k}}R\right)e_j\left(\mathbf{\hat{k}'}R\right)
    \end{equation}
    \begin{equation}
        G_{LR}(\mathbf{k},\mathbf{k'}) = \left[ \epsilon_\mu - \epsilon_m + \tilde{\epsilon}_Qkk'  \hat{k}_j\hat{k}'_j\right]e_i^*\left(\mathbf{\hat{k}}L\right)e_i\left(\mathbf{\hat{k}'}R\right)  + \tilde{\epsilon}_Qkk' \hat{k}_j \hat{k}'_ie_i^*\left(\mathbf{\hat{k}}L\right)e_j\left(\mathbf{\hat{k}'}R\right)
    \end{equation}
    \begin{equation}
        G_{RL}(\mathbf{k},\mathbf{k'}) = \left[ \epsilon_\mu - \epsilon_m + \tilde{\epsilon}_Q kk' \hat{k}_j\hat{k}'_j\right]e_i^*\left(\mathbf{\hat{k}}R\right)e_i\left(\mathbf{\hat{k}'}L\right) + \tilde{\epsilon}_Qkk' \hat{k}_j \hat{k}'_ie_i^*\left(\mathbf{\hat{k}}R\right)e_j\left(\mathbf{\hat{k}'}L\right).
    \end{equation}
\end{subequations}
\end{widetext}
In Eq. (\ref{eq_Gdefs}) we have introduced the unitless relative electric dipole strength:
\begin{equation}
    \epsilon_\mu \equiv \frac{\left(\mu^{01}\right)^2}{D^{01}}, 
\end{equation}
the unitless relative magnetic dipole strength:
\begin{equation}
    \epsilon_m \equiv \frac{\left(m^{01}\right)^2}{c^2 D^{01}}, 
\end{equation}
the unitless dissymmetry factor \cite{barron2004molecular}:
\begin{equation}
    \mathscr{G} \equiv \frac{4\mathscr{R}^{01}}{c D^{01}},
\end{equation}
as well as the parameter
\begin{equation}
    \tilde{\epsilon}_Q \equiv \frac{ \mathcal{Q}\left(\mathbf{Q^{01}}\right)}{D^{01}},
\end{equation}
which has units [length]$^2$. The unitless relative electric quadrupole strength can be defined:
\begin{equation}
    \epsilon_Q \equiv \tilde{\epsilon}_Q\frac{\sum_\mathbf{k}k^3\left|\Phi(k)\right|^2}{\sum_\mathbf{k}k\left|\Phi(k)\right|^2},
\end{equation}
or, upon taking the continuum limit $\sum_\mathbf{k} \to \frac{V}{(2\pi)^3}\int\mathrm{d}^3\mathbf{k}$:
\begin{equation}
    \epsilon_Q = \tilde{\epsilon}_Q\frac{\int_0^\infty\mathrm{d}k \, k^5 \left|\Phi(k)\right|^2}{\int_0^\infty\mathrm{d}k \, k^3 \left|\Phi(k)\right|^2} \approx \tilde{\epsilon}_Q k_0^2 \equiv \epsilon_Q^0,
\end{equation}
where the approximate equality on the right holds in the case $\left|\Phi(k)\right|^2 \to \delta(k-k_0)$. The definitions above imply
\begin{equation}
    \epsilon_\mu + \epsilon_m + \epsilon_Q = 1.
\end{equation}
Note we can alternatively write $\mathscr{G}$:
\begin{equation} \label{eq_dissymmetryvstheta}
    \mathscr{G} = 4\sqrt{\epsilon_\mu \epsilon_m} \cos\theta_{\mu m},
\end{equation}
where $\theta_{\mu m}$ is the angle between the transition electric and magnetic dipoles.
\subsection{Block diagonal form of the full density matrix}
For much of our analysis it will be useful to have a more compact representation of the density operator described by Eqs. (\ref{eq_rhopmusefuldefn}) and (\ref{eq_Gdefs}). For $s\in\{L,R\}$ and $i,j\in\{x,y,z\}$, define the unnormalized state vectors:
\begin{equation}
    \ket{\tilde{v}_i s} \equiv \frac{\sum_\mathbf{k} \Phi(k) \sqrt{k}e^*_i\left(\mathbf{\hat{k}}s\right)\ket{1(\mathbf{k}s)}}{\sqrt{2 \left(\sum_\mathbf{k}k\left|\Phi(k)\right|^2\right)}}
\end{equation}
and
\begin{equation}
    \ket{\tilde{u}_{ij}s} = \frac{\sum_\mathbf{k} \Phi(k) \sqrt{k} k_j e^*_i\left(\mathbf{\hat{k}}s\right)\ket{1(\mathbf{k}s)}}{\sqrt{2 \left(\sum_\mathbf{k}k\left|\Phi(k)\right|^2\right)}}.
\end{equation}
In terms of these vectors, $\bar{\rho}_\pm$ can be rewritten:
\begin{eqnarray} \label{eq_rhowithtildevectors}
    \bar{\rho}_\pm &=& \left(\epsilon_\mu \mp \frac{\mathscr{G}}{2} + \epsilon_m\right) \ket{\tilde{v}_iL}\bra{\tilde{v}_iL} \nonumber \\ &+& \left(\epsilon_\mu \pm \frac{\mathscr{G}}{2} + \epsilon_m\right) \ket{\tilde{v}_iR}\bra{\tilde{v}_iR} \nonumber \\ &+& (\epsilon_\mu-\epsilon_m)\bigg( \ket{\tilde{v}_iL}\bra{\tilde{v}_iR} + \ket{\tilde{v}_iR}\bra{\tilde{v}_iL} \bigg) \nonumber \\ &+& \tilde{\epsilon}_Q \Bigg\{ \ket{\tilde{u}_{ij}L}\Big[\bra{\tilde{u}_{ij}L} + \bra{\tilde{u}_{ji}L}\Big] \nonumber \\ &+& \ket{\tilde{u}_{ij}R}\Big[\bra{\tilde{u}_{ij}R} + \bra{\tilde{u}_{ji}R}\Big] \nonumber \\ &+& \ket{\tilde{u}_{ij}L}\Big[\bra{\tilde{u}_{ij}R} + \bra{\tilde{u}_{ji}R}\Big] \nonumber \\ &+& \ket{\tilde{u}_{ij}R}\Big[\bra{\tilde{u}_{ij}L} + \bra{\tilde{u}_{ji}L}\Big] \Bigg\},
\end{eqnarray}
from which it is clear that $\bar{\rho}_\pm$ is finite and at most rank-24 (= $2\times3 + 2\times3\times3$). Consider the inner product $\braket{\tilde{v}_i s|\tilde{v}_{i'}s'}$:
\begin{equation}
    \braket{\tilde{v}_i s|\tilde{v}_{i'}s'} = \frac{\delta_{ss'} \sum_\mathbf{k} k\left|\Phi(k)\right|^2 e_i\left(\mathbf{\hat{k}}s\right)e^*_{i'}\left(\mathbf{\hat{k}}s\right)}{2\left(\sum_\mathbf{k}k \left|\Phi(k)\right|^2\right)}.
\end{equation}
Taking the continuum limit then gives:
\begin{eqnarray}
    \braket{\tilde{v}_i s|\tilde{v}_{i'}s'} &=& \frac{\delta_{ss'}}{8\pi} \int \mathrm{d}\Omega_\mathbf{k} e_i\left(\mathbf{\hat{k}}s\right)e_{i'}^*\left(\mathbf{\hat{k}}s\right) \nonumber \\ &=& \frac{\delta_{ss'}\delta_{ii'}}{6}
\end{eqnarray}
where the last line follows from a relation given in Appendix \ref{appendix_wavevectorintegrals}. Hence, upon defining:
\begin{equation}
    \ket{v_i s} \equiv \sqrt{6} \ket{\tilde{v}_i s},
\end{equation}
it is evident that 
\begin{eqnarray}
\mathscr{B}^\parallel &&= \left\{\ket{v_x L},\ket{v_x R},\ket{v_y L}\ket{v_y R},\ket{v_z L},\ket{v_z R}\right\} \nonumber \\ &&\equiv \{\ket{v_1},...,\ket{v_6}\}
\end{eqnarray}
forms an (ordered) orthonormal basis for the subspace spanned by these vectors. 

Now consider an inner product of the form
\begin{equation}
    \Big[ \bra{\tilde{u}_{ij}s} + \bra{\tilde{u}_{ji}s} \Big] \ket{\tilde{v}_{i'}s'} \propto \varepsilon_{i'ij} + \varepsilon_{i'ji} = 0,
\end{equation}
where once again we've made use of the relations in Appendix \ref{appendix_wavevectorintegrals}. Hence we can conclude that $\text{span}(\mathscr{B}^\parallel)$ is orthogonal to the operator contained within the curly brackets in Eq. (\ref{eq_rhowithtildevectors}). A complementary orthonormal basis $\mathscr{B}^\perp = \{\ket{u_7},...,\ket{u_{\text{rank}(\bar{\rho}_\pm)}}\}$ can be constructed by continuing the Gram-Schmidt process with the vectors $\{\ket{\tilde{u}_{ij}s}\}$ such that $\text{span}(\mathscr{B}^\parallel\cup\mathscr{B}^\perp)=\text{im}(\bar{\rho}_\pm)$ and $\braket{v|u}=0 \, \forall \, \ket{v} \in \text{span}(\mathscr{B}^\parallel), \ket{u} \in \text{span}(\mathscr{B}^\perp)$. In the constructed basis, $\bar{\rho}_\pm$ is represented by a block diagonal matrix of the form:
\begin{equation} \label{eq_rhomatrix}
    \boldsymbol{\bar{\uprho}_\pm} =
        \begin{pmatrix}
  \boldsymbol{\bar{\uprho}_\pm^{(\parallel)}} & \mathbf{0} \\  
  \mathbf{0} & \boldsymbol{\bar{\uprho}^{(\mathscr{\perp})}} 
\end{pmatrix}
\end{equation}
where the constituent submatrices have elements:
\begin{subequations}
    \begin{equation}
        \left[ \boldsymbol{\bar{\uprho}_\pm^{(\parallel)}} \right]_{ll'} = \braket{v_l|\bar{\rho}_\pm|v_{l'}}, \, \forall \ket{v_l}, \ket{v_{l'}} \in \mathscr{B}^\parallel
    \end{equation}
    \begin{equation}
        \left[ \boldsymbol{\bar{\uprho}^{(\perp)}} \right]_{ll'} = \braket{u_{l+6}|\bar{\rho}_\pm|u_{l'+6}}, \, \forall \ket{u_{l+6}}, \ket{u_{l'+6}} \in \mathscr{B}^\perp.
    \end{equation}
\end{subequations}
Importantly, the submatrix $\boldsymbol{\bar{\uprho}^{(\perp)}}$ is the same for both stereoisomers and its trace is simply:
\begin{equation}
    \text{Tr}\boldsymbol{\bar{\uprho}^{(\perp)}} = \epsilon_Q
\end{equation}
For the remainder of the analysis we shall be content to leave $\boldsymbol{\bar{\uprho}^{(\perp)}}$ in generic form without specifying its elements. The elements of $\boldsymbol{\bar{\uprho}_\pm^{(\parallel)}}$ can be computed with the help of relations given in Appendix \ref{appendix_wavevectorintegrals} to yield:
\begin{equation}
    \boldsymbol{\bar{\uprho}_\pm^{(\parallel)}} = \left(\epsilon_\mu + \epsilon_m\right)\frac{\mathbf{I_3}}{3} \otimes \varrho_\pm,
\end{equation}
where $\mathbf{I_3}$ is the $3\times3$ identity matrix and we've defined the normalized qubit states:
\begin{equation} \label{eq_qubitstates}
    \varrho_\pm \equiv \frac{1}{2}\left(I + \tilde{\chi}\sigma_x \mp \frac{\tilde{\mathscr{G}}}{2} \sigma_z\right),
\end{equation}
with:
\begin{subequations} \label{eq_chianddeltatilde}
    \begin{equation}
        \tilde{\chi} \equiv \frac{\epsilon_\mu-\epsilon_m}{\epsilon_\mu+\epsilon_m} \in [-1,1],
    \end{equation}
    \begin{equation}
        \tilde{\mathscr{G}} \equiv \frac{\mathscr{G}}{\epsilon_\mu+\epsilon_m},
    \end{equation}
\end{subequations}
and where $I$ is the identity, $\sigma_x$ is the Pauli-$x$ operator, and $\sigma_z$ is the Pauli-$z$ operator. The form of $\boldsymbol{\bar{\uprho}_\pm}$ given in Eq. (\ref{eq_rhomatrix}) could have presumably been arrived at by starting the problem by expressing $\boldsymbol{\upmu^{01}}$, $\mathbf{m^{01}}$, and $\mathbf{Q^{01}}$ in terms of spherical (rather than Cartesian) tensor components, and using the less-commonly encountered multipole expansions of the quantum field operators \cite{heitler1954quantum,Shumovsky2001} in place of the plane-wave expansions given in Eq. (\ref{eq_fieldoperators}). The block-diagonal form arises from group theoretical considerations that are implicit in the tensorial relations listed in Appendix \ref{appendix_rotationaavs}. However, also having the representation of $\bar{\rho}_\pm$ given in Eqs. (\ref{eq_rhopmusefuldefn}) and (\ref{eq_Gdefs}) will prove useful when we consider the performance of accessible measurement strategies.
\subsection{Isolating multipolar parameters with angle-resolved coherent detection}
\begin{figure*}
    \centering
    \includegraphics[width=\linewidth]{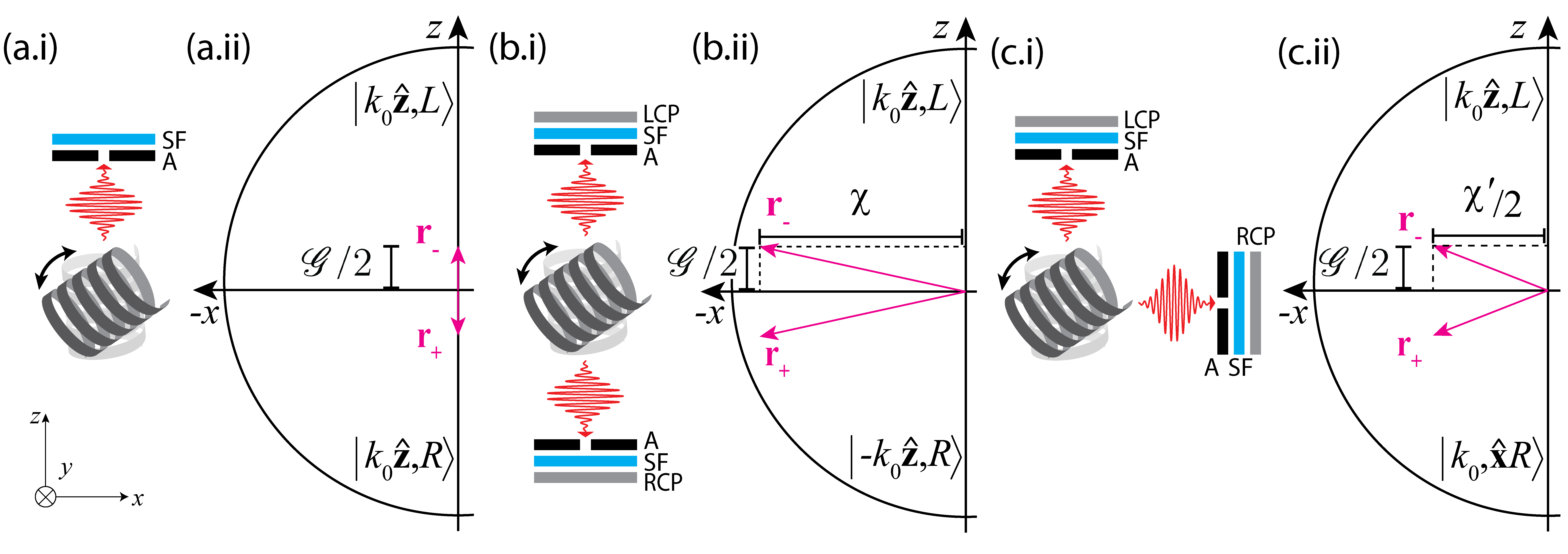}
    \caption{Filtered collection from one (a.i) or two (b.i), (c.i) directions. A: aperture, SF: spectral filter, LCP/RCP: left/right-circular polarizer. (a.ii), (b.ii), and (c.ii) sketch the Bloch vectors $\mathbf{r}_\pm$ for the qubit states described in the main text, assuming $\chi, \chi', \mathscr{G}>0$.}
    \label{fig:collectionschemes}
\end{figure*}
To simplify the remainder of our analysis, we next imagine post-selecting the photon state given by Eqs. (\ref{eq_rhopmusefuldefn}) and (\ref{eq_Gdefs}) on a single wave number $k_0$ (which can be approximately achieved by collecting through a narrow spectral filter) and on one or two particular propagation directions $\mathbf{\hat{k}}$ (which can be approximately achieved by collecting through a narrow aperture) (Fig. \ref{fig:collectionschemes}).

\begin{figure}
    \centering
    \includegraphics[width=\linewidth]{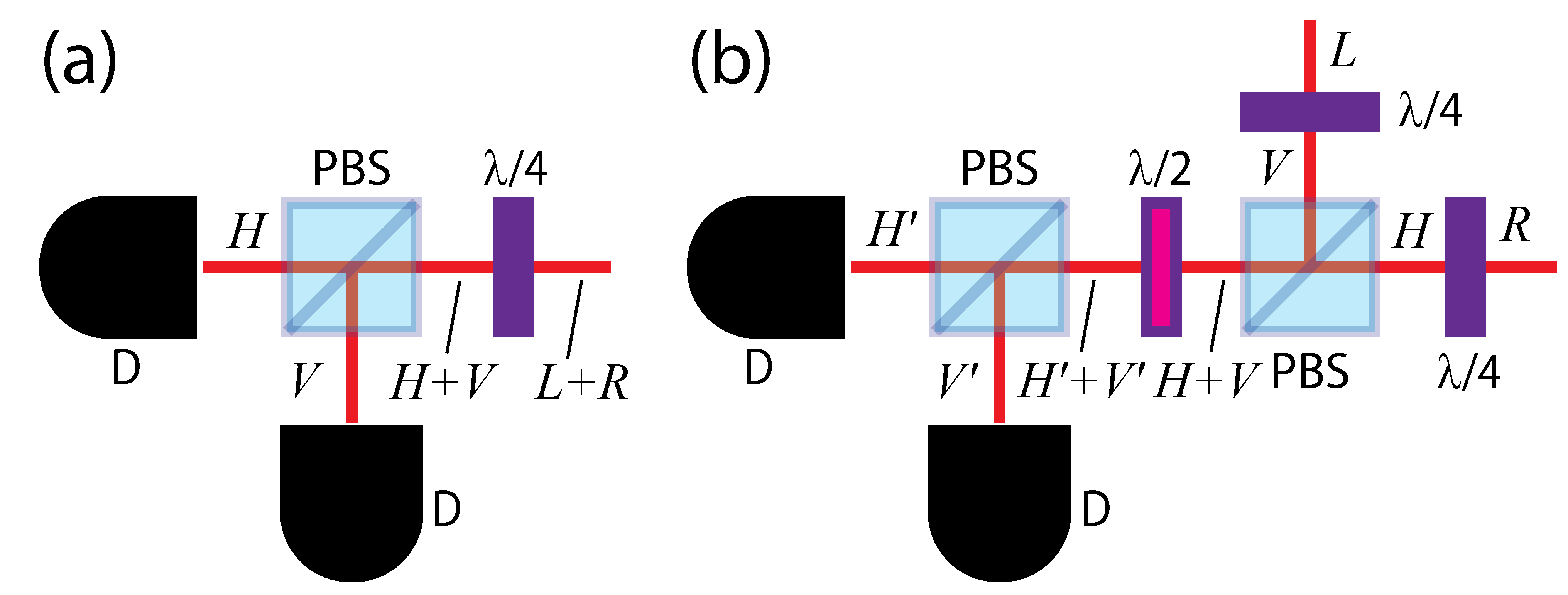}
    \caption{Detection schemes. (a) The conventional approach to chiroptical detection. A single spatial mode containing both $L$- and $R$-polarized light is transmitted through a quarter wave plate ($\lambda/4$), converting one circular polarization to horizontally polarized light ($H$) and the other to vertically polarized light ($V$). The polarization channels are split by a polarizing beam splitter (PBS) and photons are counted on the detectors (D). (b) Coherent detection of chiroptical emission collected from two different spatial modes, as in Fig. \ref{fig:collectionschemes}b, c. $L$-polarized light from one spatial mode and $R$-polarized light from another are converted to $H$ and $V$ with quarter wave plates then coherently combined into a single spatial mode with a polarizing beam splitter. A half wave plate ($\lambda/2$) rotates the plane of linear polarization, then photons are counted in the rotated basis.}
    \label{fig:coherentdetection}
\end{figure}

For one such example, suppose we post-select the state of the photon given by Eqs. (\ref{eq_rhopmusefuldefn}) and (\ref{eq_Gdefs}) to permit only a single $\mathbf{k}$, say with wavenumber $k_0$ propagating in the $+\hat{\mathbf{z}}$ direction (Fig. \ref{fig:collectionschemes}a.i). The normalized state of a collected photon is then:
\begin{eqnarray}
    \bar{\rho}_\pm^{(\hat{\mathbf{z}})} = \frac{1}{2}\left[\epsilon_\mu \mp \frac{\mathscr{G}}{2} + \epsilon_m + \epsilon_Q^0 \right]\ket{1(k_0\hat{\mathbf{z}},L)}\bra{1(k_0\hat{\mathbf{z}},L)} \nonumber \\ + \frac{1}{2}\left[\epsilon_\mu \pm \frac{\mathscr{G}}{2} + \epsilon_m + \epsilon_Q^0 \right]\ket{1(k_0\hat{\mathbf{z}},R)}\bra{1(k_0\hat{\mathbf{z}},R)} \nonumber \\ = \frac{1}{2}\left[1 \mp \frac{\mathscr{G}}{2}\right]\ket{1(k_0\hat{\mathbf{z}},L)}\bra{1(k_0\hat{\mathbf{z}},L)} \nonumber \\ + \frac{1}{2}\left[1 \pm \frac{\mathscr{G}}{2} \right]\ket{1(k_0\hat{\mathbf{z}},R)}\bra{1(k_0\hat{\mathbf{z}},R)}
\end{eqnarray}
where we've made use of the fact that
\begin{equation}
    \epsilon_\mu + \epsilon_m + \epsilon_Q^0=1
\end{equation} 
in this case. Equivalently, in the basis $\left\{\ket{1(k_0\hat{\mathbf{z}},L)},\ket{1(k_0\hat{\mathbf{z}},R)}\right\}$ we can write:
\begin{equation}
    \bar{\rho}_\pm^{(\hat{\mathbf{z}})} = \frac{1}{2}\left( I \mp \frac{\mathscr{G}}{2}\sigma_z \right).
\end{equation}
The Bloch vectors for these qubit states, which are relevant for the typical approach to measuring CPL in which light is collected from a single direction, are sketched in Fig. \ref{fig:collectionschemes}a.ii. Using a polarizing beam splitter and two photon counters (Fig. \ref{fig:coherentdetection}a) one can effectively measure $\sigma_z$ and thus deduce $\mathscr{G}$. Upon inspection of Eq. (\ref{eq_dissymmetryvstheta}) it is clear that $\mathscr{G}$ conflates three distinct physical parameters in $\epsilon_\mu$, $\epsilon_m$, and $\theta_{\mu m}$. Evidently, the individual contributions of these three parameters cannot be separately determined from a unidirectional measurement of an unoriented sample in the absence of external fields.

More information can be recovered by collecting from at least one additional direction. For our next example, we post-select the state of the photon given by Eqs. (\ref{eq_rhopmusefuldefn}) and (\ref{eq_Gdefs}) to permit only a single $\mathbf{k}$ and its opposite, say with wavenumber $k_0$ propagating in the $+\hat{\mathbf{z}}$ or $-\hat{\mathbf{z}}$ directions (Fig. \ref{fig:collectionschemes}b.i). The state of a collected photon is then:
\begin{widetext}
\begin{eqnarray}
 \bar{\rho}_\pm^{(\hat{\mathbf{z}},-\hat{\mathbf{z}})} = \frac{1}{4}\left[1 \mp \frac{\mathscr{G}}{2} \right]\bigg[\ket{1(k_0\hat{\mathbf{z}},L)}\bra{1(k_0\hat{\mathbf{z}},L)}+\ket{1(-k_0\hat{\mathbf{z}},L)}\bra{1(-k_0\hat{\mathbf{z}},L)}\bigg] \nonumber \\ + \frac{1}{4}\left[1 \pm \frac{\mathscr{G}}{2} \right]\bigg[\ket{1(k_0\hat{\mathbf{z}},R)}\bra{1(k_0\hat{\mathbf{z}},R)}+\ket{1(-k_0\hat{\mathbf{z}},R)}\bra{1(-k_0\hat{\mathbf{z}},R)}\bigg] \nonumber \\ - \frac{1}{4}\left[\epsilon_\mu - \epsilon_m - \epsilon_Q^0\right]\bigg[\ket{1(k_0\hat{\mathbf{z}},L)}\bra{1(-k_0\hat{\mathbf{z}},R)}+\ket{1(-k_0\hat{\mathbf{z}},R)}\bra{1(k_0\hat{\mathbf{z}},L)}\bigg] \nonumber \\ - \frac{1}{4}\left[\epsilon_\mu - \epsilon_m -\epsilon_Q^0 \right]\bigg[\ket{1(k_0\hat{\mathbf{z}},R)}\bra{1(-k_0\hat{\mathbf{z}},L)}+\ket{1(-k_0\hat{\mathbf{z}},L)}\bra{1(k_0\hat{\mathbf{z}},R)}\bigg]. \nonumber \\
\end{eqnarray}
\end{widetext}
The presence of coherence between modes of opposite polarizations propagating in opposite directions is significant to our findings. As shown in Appendix \ref{appendix_classicalED}, this coherence can be understood equally well from a classical electrodynamics treatment. 

In the basis $\{\ket{1(k_0\hat{\mathbf{z}},L)}$, $\ket{1(-k_0\hat{\mathbf{z}},R)}$, $\ket{1(-k_0\hat{\mathbf{z}},L)}$, $\ket{1(k_0\hat{\mathbf{z}},R)}\}$ (note the alternating order) the state can be represented succinctly:
\begin{equation}
    \bar{\rho}_\pm^{(\hat{\mathbf{z}},-\hat{\mathbf{z}})} = \frac{I}{4} \otimes \left( I - \chi\sigma_x\mp \frac{\mathscr{G}}{2} \sigma_z\right),
\end{equation}
where we've introduced yet another unitless constant:
\begin{equation} \label{eq_chidefn}
    \chi \equiv \epsilon_\mu-\epsilon_m-\epsilon_Q^0\in[-1,1].
\end{equation}
For our purposes we can simplify even further by assuming that we insert an $L$-polarizer in the path of the $+\mathbf{\hat{z}}$ mode and an $R$-polarizer in the path of the $-\mathbf{\hat{z}}$ mode. The resulting qubit states are:
\begin{equation} \label{eq_rhopmz}
    \bar{\rho}_\pm^{(\mathbf{\hat{z}}L,-\mathbf{\hat{z}}R)} = \frac{1}{2}\left(I -\chi\sigma_x \mp \frac{\mathscr{G}}{2} \sigma_z\right),
\end{equation}
the Bloch vectors of which are sketched in Fig. \ref{fig:collectionschemes}b.ii. Clearly, $\chi$ can be deduced from a measurement of $\sigma_x$, as realized by the appropriate sequence of polarizing beam splitters and wave plates (Fig. \ref{fig:coherentdetection}b).

As our third and final example of this type, we consider filtering such that light of a single wavenumber $k_0$ propagating in two perpendicular directions, $+\mathbf{\hat{z}}$ and $+\mathbf{\hat{x}}$, is collected (Fig. \ref{fig:collectionschemes}c.i).  The state of the photon is:
\begin{widetext}
\begin{eqnarray}
 \bar{\rho}_\pm^{(\hat{\mathbf{z}},\hat{\mathbf{x}})} &=& \frac{1}{4}\left[1 \mp \frac{\mathscr{G}}{2} \right]\bigg[\ket{1(k_0\hat{\mathbf{z}},L)}\bra{1(k_0\hat{\mathbf{z}},L)}+\ket{1(k_0\hat{\mathbf{x}},L)}\bra{1(k_0\hat{\mathbf{x}},L)}\bigg] \nonumber \\ &+& \frac{1}{4}\left[1 \pm \frac{\mathscr{G}}{2} \right]\bigg[\ket{1(k_0\hat{\mathbf{z}},R)}\bra{1(k_0\hat{\mathbf{z}},R)}+\ket{1(k_0\hat{\mathbf{x}},R)}\bra{1(k_0\hat{\mathbf{x}},R)}\bigg] \nonumber \\ &+& \frac{1}{8}\left[\epsilon_\mu + \epsilon_m - \epsilon_Q^0 \mp \frac{\mathscr{G}}{2}\right]\bigg[\ket{1(k_0\hat{\mathbf{z}},L)}\bra{1(k_0\hat{\mathbf{x}},L)}+\ket{1(k_0\hat{\mathbf{x}},L)}\bra{1(k_0\hat{\mathbf{z}},L)}\bigg] \nonumber \\ &+& \frac{1}{8}\left[\epsilon_\mu + \epsilon_m - \epsilon_Q^0 \pm \frac{\mathscr{G}}{2}\right]\bigg[\ket{1(k_0\hat{\mathbf{z}},R)}\bra{1(k_0\hat{\mathbf{x}},R)}+\ket{1(k_0\hat{\mathbf{x}},R)}\bra{1(k_0\hat{\mathbf{z}},R)}\bigg] \nonumber \\ &-& \frac{1}{8}\left[\epsilon_\mu - \epsilon_m + \epsilon_Q^0 \right]\bigg[\ket{1(k_0\hat{\mathbf{z}},L)}\bra{1(k_0\hat{\mathbf{x}},R)}+\ket{1(k_0\hat{\mathbf{x}},R)}\bra{1(k_0\hat{\mathbf{z}},L)}\bigg] \nonumber \\ &-& \frac{1}{8}\left[\epsilon_\mu - \epsilon_m + \epsilon_Q^0\right]\bigg[\ket{1(k_0\hat{\mathbf{z}},R)}\bra{1(k_0\hat{\mathbf{x}},L)}+\ket{1(k_0\hat{\mathbf{x}},L)}\bra{1(k_0\hat{\mathbf{z}},R)}\bigg], \nonumber \\
\end{eqnarray}
\end{widetext}
which, in the basis $\{\ket{1(k_0\hat{\mathbf{z}},L)}$, $\ket{1(k_0\hat{\mathbf{x}},R)}$, $\ket{1(k_0\hat{\mathbf{x}},L)}$, $\ket{1(k_0\hat{\mathbf{z}},R)}\}$, can be written more succinctly as:
\begin{eqnarray}
    \bar{\rho}_\pm^{(\mathbf{\hat{z}},\mathbf{\hat{x}})} = && \frac{I}{4}\otimes\left(I - \frac{\chi'}{2}\sigma_x \mp \frac{\mathscr{G}}{2} \sigma_z\right) \nonumber \\ &&+\frac{\sigma_x}{8}\otimes\left(\chi''I \mp \frac{\mathscr{G}}{2}\sigma_z\right),
\end{eqnarray}
where we've introduced two new constants:
\begin{subequations}
    \begin{equation}
        \chi' \equiv \epsilon_\mu - \epsilon_m + \epsilon_Q^0 \in [-1,1],
    \end{equation}
    \begin{equation}
        \chi'' \equiv \epsilon_\mu + \epsilon_m - \epsilon_Q^0 \in [-1,1].
    \end{equation}
\end{subequations}
Once more, for our purposes we can simplify even further by assuming that we insert an $L$-polarizer in the path of the $\mathbf{\hat{z}}$ mode and an $R$-polarizer in the path of the $\mathbf{\hat{x}}$ mode. The resulting qubit states are:
\begin{equation}
    \bar{\rho}_\pm^{(\mathbf{\hat{z}}L,\mathbf{\hat{x}}R)} = \frac{1}{2}\left(I - \frac{\chi'}{2}\sigma_x \mp\frac{\mathscr{G}}{2}\sigma_z\right),
\end{equation}
the Bloch vectors of which are sketched in Fig. \ref{fig:collectionschemes}c.ii. Similar to the previous case, $\chi'$ can be deduced from a measurement of $\sigma_x$, as realized by the appropriate sequence of polarizing beam splitters and wave plates (Fig. \ref{fig:coherentdetection}b).
Altogether, the collection and detection schemes depicted in Figs. \ref{fig:collectionschemes} and \ref{fig:coherentdetection} yield estimates of $\mathscr{G}$, $\chi$, and $\chi'$. Algebraic combinations of these three parameters in turn can generate estimates of the unknown physical parameters $\epsilon_\mu$, $\epsilon_m$, and $\theta_{\mu m}$. Thus, by implementing the right sequence of coherent measurements, one is able to tease apart distinct physical contributions to chiral light-matter interactions that would otherwise be unresolved by implementing the conventional scheme (Figs. \ref{fig:collectionschemes}a and \ref{fig:coherentdetection}a) alone.

\section{Conclusion}
In conclusion, we have derived the state of the quantum electromagnetic field resulting from spontaneous emission of a randomly oriented quantum emitter, including the first three multipole moments (electric dipole, magnetic dipole, and electric quadrupole). This sets the stage for the analysis of the quantum limits to chiroptical discrimination presented in Ref. \cite{Backlund2026Letter}. Here we propose a sequence of single-photon measurement arrangements in which light emitted in opposite polarizations and different directions is coherently recombined and detected in order to disambiguate the various molecular parameters that contribute to chiroptical interactions, even for an unoriented sample.

\begin{acknowledgements}
    This work was supported by National Science Foundation award number 2441430. Thanks to Chris Anderson for illuminating discussion.
\end{acknowledgements}

\appendix
\section{Rotational tensor averages} \label{appendix_rotationaavs}
This Appendix is a concise summary of useful results from Appendix 2 of Ref. \cite{craig1998molecular}. The components of a general tensor $\mathbf{T}$ of rank $n$ expressed in one choice of coordinates obey the transformation law:
\begin{equation}
    T'_{i_1 \cdots i_n} = R_{i_1\lambda_1}\cdots R_{i_n\lambda_n} T_{\lambda_1 \cdots \lambda_n},
\end{equation}
where Greek indices label the old coordinate system, Latin indices label the new coordinate system, and $R_{i\lambda}$ is the direction cosine connecting the indicated old and new axes. We denote the rotational average of products of direction cosines as:
\begin{eqnarray}
    I^{(n)} &=& \langle R_{i_1\lambda_1}\cdots R_{i_n\lambda_n} \rangle \nonumber \\ &\equiv& \frac{1}{8 \pi^2} \int_0^{2\pi}\mathrm{d}\psi\int_0^{\pi}\mathrm{d}\theta \sin\theta \int_0^{2\pi}\mathrm{d}\phi \, R_{i_1\lambda_1}\cdots R_{i_n\lambda_n}. \nonumber \\
\end{eqnarray}
Below we summarize results for $n=1$ through $n=4$.
\begin{subequations}
    \begin{equation}
        I^{(1)} = 0
    \end{equation}
    \begin{equation}
        I^{(2)} = \frac{1}{3} \delta_{i_1 i_2} \delta_{\lambda_1\lambda_2}
    \end{equation}
    \begin{equation}
        I^{(3)} = \frac{1}{6} \varepsilon_{i_1i_2i_3} \varepsilon_{\lambda_1 \lambda_2 \lambda_3}
    \end{equation}
    \begin{equation}
        I^{(4)} = \frac{1}{30} \begin{pmatrix}
            \delta_{i_1 i_2} \delta_{i_3 i_4} \\ \delta_{i_1 i_3} \delta_{i_2 i_4} \\ \delta_{i_1 i_4} \delta_{i_2 i_3}
        \end{pmatrix}^\text{T} \begin{pmatrix}
            4 & -1 & -1 \\ -1 & 4 & -1 \\ -1 & -1 & 4
        \end{pmatrix} \begin{pmatrix}
            \delta_{\lambda_1\lambda_2} \delta_{\lambda_3\lambda_4} \\ \delta_{\lambda_1\lambda_3} \delta_{\lambda_2\lambda_4} \\ \delta_{\lambda_1\lambda_4} \delta_{\lambda_2\lambda_3}
        \end{pmatrix}.
    \end{equation}
\end{subequations}
\section{Angular integrals of wave-vector products} \label{appendix_wavevectorintegrals}
Angular integrals of the form $\int \mathrm{d}\Omega_\mathbf{k} \, \hat{k}_{i_1}\cdots\hat{k}_{i_n}$ can be evaluated with the help of the identities summarized in Appendix \ref{appendix_rotationaavs}. In particular, an arbitrary $\mathbf{\hat{k}}$ can be written as a rotation of the unit vector $\mathbf{\hat{z}}$. In coordinate form:
\begin{equation}
    \hat{k}_i = R_{i\lambda} \hat{z}_\lambda = R_{i\lambda}\delta_{\lambda 3}.
\end{equation}
Hence,
\begin{eqnarray}
    \langle \hat{k}_{i_1} \cdots \hat{k}_{i_n} \rangle &=& \langle R_{i_1\lambda_1}\cdots R_{i_n\lambda_n} \rangle \delta_{\lambda_1 3}\cdots\delta_{\lambda_n 3} \nonumber \\ &=& I^{(n)} \delta_{\lambda_1 3} \cdots \delta_{\lambda_n 3}.
\end{eqnarray}
Let $\theta_\mathbf{k},\phi_\mathbf{k}$ be the polar and azimuthal angles of $\mathbf{\hat{k}}$, respectively, defined with respect to the lab-fixed frame. Let $\psi_\mathbf{k}$ be the angle defined about $\mathbf{k}$. We have:
\begin{eqnarray}
    \langle \hat{k}_{i_1} &\cdots& \hat{k}_{i_n} \rangle \nonumber \\ &&= \frac{1}{8 \pi^2} \int_0^{2\pi}\mathrm{d}\psi_\mathbf{k}\int_0^{\pi}\mathrm{d}\theta_\mathbf{k} \sin\theta_\mathbf{k} \int_0^{2\pi}\mathrm{d}\phi_\mathbf{k} \, \hat{k}_{i_1}\cdots \hat{k}_{i_n} \nonumber \\ &&= \frac{1}{4\pi} \int_0^{\pi}\mathrm{d}\theta_\mathbf{k} \sin\theta_\mathbf{k} \int_0^{2\pi}\mathrm{d}\phi_\mathbf{k} \, \hat{k}_{i_1}\cdots \hat{k}_{i_n} \nonumber \\ &&= \frac{1}{4\pi} \int \mathrm{d}\Omega_\mathbf{k} \, \hat{k}_{i_1}\cdots \hat{k}_{i_n},
\end{eqnarray}
where the second line follows from the invariance of $\mathbf{\hat{k}}$ upon rotation about $\mathbf{\hat{k}}$. Thus we conclude:
\begin{equation}
    \int \mathrm{d}\Omega_\mathbf{k} \, \hat{k}_{i_1}\cdots \hat{k}_{i_n} = 4\pi I^{(n)} \delta_{\lambda_1 3} \cdots \delta_{\lambda_n 3}.
\end{equation}
For $n=1$ through $n=4$ this gives:
\begin{subequations}
    \begin{equation}
        \int \mathrm{d}\Omega_\mathbf{k} \, \hat{k}_{i_1} = 0
    \end{equation}
    \begin{equation}
        \int \mathrm{d}\Omega_\mathbf{k} \, \hat{k}_{i_1}\hat{k}_{i_2} = \frac{4\pi}{3}\delta_{i_1 i_2}
    \end{equation}
    \begin{equation}
        \int \mathrm{d}\Omega_\mathbf{k} \, \hat{k}_{i_1}\hat{k}_{i_2}\hat{k}_{i_3} = 0
    \end{equation}
    \begin{eqnarray}
        \int \mathrm{d}\Omega_\mathbf{k} && \, \hat{k}_{i_1}\hat{k}_{i_2}\hat{k}_{i_3}\hat{k}_{i_4} = \nonumber \\ &&\frac{8\pi}{30} \left( \delta_{i_1 i_2}\delta_{i_3 i_4} + \delta_{i_1 i_3}\delta_{i_2 i_4} + \delta_{i_1 i_4}\delta_{i_2 i_3} \right). \nonumber \\
    \end{eqnarray}
\end{subequations}
In our analysis we also require computation of angular integrals of the form $\int\mathrm{d}\Omega_\mathbf{k} e^*_i\left(\mathbf{\hat{k}}L\right)e_{i'}\left(\mathbf{\hat{k}}L\right)$, $\int\mathrm{d}\Omega_\mathbf{k} \hat{k}_je^*_i\left(\mathbf{\hat{k}}L\right)e_{i'}\left(\mathbf{\hat{k}}L\right)$, and $\int\mathrm{d}\Omega_\mathbf{k} \hat{k}_j\hat{k}_{j'}e^*_i\left(\mathbf{\hat{k}}L\right)e_{i'}\left(\mathbf{\hat{k}}L\right)$, plus the analogous integrals with $L\to R$. The vectors $\mathbf{e}\left(\mathbf{\hat{k}}L\right), \mathbf{e}\left(\mathbf{\hat{k}}R\right)$ can be obtained by a 3D rotation of $(\mathbf{\hat{x}}+i\mathbf{\hat{y}})/\sqrt{2}$ and $(\mathbf{\hat{x}}-i\mathbf{\hat{y}})/\sqrt{2}$, respectively
\begin{subequations}
    \begin{equation}
        e_i(\mathbf{\hat{k}}L) = R_{i\lambda}
        \frac{\left(\hat{x}_\lambda+i\hat{y}_\lambda\right)}{\sqrt{2}} = R_{i\lambda}\frac{\left(\delta_{\lambda 1}+i\delta_{\lambda2}\right)}{\sqrt{2}}.
    \end{equation}
    \begin{equation}
        e_i(\mathbf{\hat{k}}R) = R_{i\lambda}
        \frac{\left(\hat{x}_\lambda-i\hat{y}_\lambda\right)}{\sqrt{2}} = R_{i\lambda}\frac{\left(\delta_{\lambda 1}-i\delta_{\lambda2}\right)}{\sqrt{2}}.
    \end{equation}
\end{subequations}
Therefore:
\begin{subequations}
    \begin{equation}
        \left\langle e_i^*\left(\mathbf{\hat{k}}L\right)e_{i'}\left(\mathbf{\hat{k}}L\right) \right\rangle = \left\langle e_i^*\left(\mathbf{\hat{k}}R\right)e_{i'}\left(\mathbf{\hat{k}}R\right) \right\rangle = \frac{\delta_{ii'}}{3}
    \end{equation}
    \begin{eqnarray}
        \left\langle e_i^*\left(\mathbf{\hat{k}}L\right)e_{i'}\left(\mathbf{\hat{k}}L\right) \hat{k}_j\right\rangle &=& \left\langle e_i^*\left(\mathbf{\hat{k}}R\right)e_{i'}\left(\mathbf{\hat{k}}R\right) \hat{k}_j\right\rangle^* \nonumber \\ &=& \frac{i \varepsilon_{ii'j}}{6}
    \end{eqnarray}
    \begin{eqnarray}
        \left\langle e_i^*\left(\mathbf{\hat{k}}L\right)e_{i'}\left(\mathbf{\hat{k}}L\right) \hat{k}_j\hat{k}_{j'}\right\rangle = \left\langle e_i^*\left(\mathbf{\hat{k}}R\right)e_{i'}\left(\mathbf{\hat{k}}R\right) \hat{k}_j\hat{k}_{j'}\right\rangle \nonumber \\ = \frac{1}{30}\left(4 \delta_{ii'}\delta_{jj'} - \delta_{ij}\delta_{i'j'} - \delta_{ij'}\delta_{i'j} \right). \nonumber \\
    \end{eqnarray}
\end{subequations}
Again taking $\psi_\mathbf{k}$ to be the angle of rotation about $\mathbf{k}$ we note that $\mathbf{e}\left(\mathbf{\hat{k}}L\right)$ [$\mathbf{e}\left(\mathbf{\hat{k}}R\right)$] only incurs a phase factor $e^{- i \psi_\mathbf{k}}$ [$e^{i\psi_\mathbf{k}}$] under rotations through $\psi_\mathbf{k}$, and therefore the products $e^*_i\left(\mathbf{\hat{k}}L\right)e_{i'}\left(\mathbf{\hat{k}}L\right)$ and $e^*_i\left(\mathbf{\hat{k}}R\right)e_{i'}\left(\mathbf{\hat{k}}R\right)$ are completely invariant under this rotation $\forall i,i'$. It follows that:
\begin{subequations}
        \begin{eqnarray} \int \mathrm{d}\Omega_\mathbf{k}\, e_i^*\left(\mathbf{\hat{k}}L\right)e_{i'}\left(\mathbf{\hat{k}}L\right) &=& \int \mathrm{d}\Omega_\mathbf{k}\,  e_i^*\left(\mathbf{\hat{k}}R\right)e_{i'}\left(\mathbf{\hat{k}}R\right) \nonumber \\ &=& \frac{4\pi\delta_{ii'}}{3}
    \end{eqnarray}
    \begin{eqnarray}
        \int \mathrm{d}\Omega_\mathbf{k} && e_i^*\left(\mathbf{\hat{k}}L\right)e_{i'}\left(\mathbf{\hat{k}}L\right) \hat{k}_j \nonumber \\ &=& \left( \int \mathrm{d}\Omega_\mathbf{k}\,e_i^*\left(\mathbf{\hat{k}}R\right)e_{i'}\left(\mathbf{\hat{k}}R\right) \hat{k}_j\right)^* \nonumber \\ &=& \frac{4\pi i \varepsilon_{ii'j}}{6}
    \end{eqnarray}
    \begin{eqnarray}
        \int \mathrm{d}\Omega_\mathbf{k}&&\, e_i^*\left(\mathbf{\hat{k}}L\right)e_{i'}\left(\mathbf{\hat{k}}L\right) \hat{k}_j\hat{k}_{j'} \nonumber \\ &=& \int \mathrm{d}\Omega_\mathbf{k}\, e_i^*\left(\mathbf{\hat{k}}R\right)e_{i'}\left(\mathbf{\hat{k}}R\right) \hat{k}_j\hat{k}_{j'} \nonumber \\ &=& \frac{4\pi}{30}\left(4 \delta_{ii'}\delta_{jj'} - \delta_{ij}\delta_{i'j'} - \delta_{ij'}\delta_{i'j} \right).
    \end{eqnarray}
\end{subequations}
\section{Spatial coherence of the classical field} \label{appendix_classicalED}
In both the main text of this article and Ref. \cite{Backlund2026Letter}, the coherences between single-photon modes of different directions and polarizations are key to our findings. While we have derived these coherences from a fully quantum mechanical perspective in the main text, here we will show that they can be understood from a classical electrodynamics approach as well. Consider a classical electric dipole $\boldsymbol{\upmu}(t)$ and magnetic dipole $\mathbf{m}(t)$ located at the origin and oscillating $\pi/2$ out of phase with angular frequency $\omega$ such that:
\begin{subequations}
    \begin{equation}
        \boldsymbol{\upmu}(t) = \boldsymbol{\upmu_0} e^{-i \omega t}
    \end{equation}
    \begin{equation}
        \mathbf{m}(t) = i\mathbf{m_0}e^{-i\omega t},
    \end{equation}
\end{subequations}
where $\boldsymbol{\upmu_0} = [\mu_x,\mu_y,\mu_z]^\text{T},\mathbf{m_0} = [m_x,m_y,m_z]^\text{T} \in \mathbb{R}^3$ and it's understood here and throughout this section that any physically observable quantity is to be equated with the real part of its complex classical wavefunction. For simplicity we will omit the electric quadrupole from this analysis. The radiated electric field at position $\mathbf{r}$ in the far field is given by \cite{jackson1999classical}:
\begin{equation}
    \mathbf{E}(\mathbf{r},t) = \frac{c k^2 Z_0}{4\pi} \frac{e^{i(kr-\omega t)}}{r} \left[ \left( \hat{\mathbf{r}} \times \boldsymbol{\upmu_0} \right)\times \hat{\mathbf{r}} - \frac{i}{c}\left( \hat{\mathbf{r}} \times \mathbf{m_0} \right) \right],
\end{equation}
where $k = \omega/c$ and $Z_0$ is the impedence of free space. For this illustrative example we will consider the mutual coherence of the field at two opposing positions in the far field $\mathbf{r_\uparrow} = [0,0,z]^\text{T}$ and $\mathbf{r_\downarrow} = [0,0,-z]^\text{T}$ for some $z>0$. Those fields are:
\begin{subequations} \label{eq_EfieldsCart}
    \begin{equation}
        \mathbf{E}(\mathbf{r_\uparrow},t) = A \left[ \left( \mu_x + \frac{i}{c}m_y \right)\hat{\mathbf{x}} + \left( \mu_y - \frac{i}{c}m_x \right)\hat{\mathbf{y}} \right],
    \end{equation}
    \begin{equation}
        \mathbf{E}(\mathbf{r_\downarrow},t) = A \left[ \left( \mu_x - \frac{i}{c}m_y \right)\hat{\mathbf{x}} + \left( \mu_y + \frac{i}{c}m_x \right)\hat{\mathbf{y}} \right],
    \end{equation}
\end{subequations}
where
\begin{equation}
    A \equiv \frac{c k^2 Z_0}{4 \pi} \frac{e^{i(kz-\omega t)}}{z}.
\end{equation}
At $\mathbf{r_\uparrow}$ the wave travels in the positive $\hat{\mathbf{z}}$ direction. An appropriately defined circular polarization basis is
\begin{subequations}
    \begin{equation}
        \mathbf{\hat{L}_\uparrow} = \frac{1}{\sqrt{2}}\left(\hat{\mathbf{x}} + i\hat{\mathbf{y}}\right)
    \end{equation}
    \begin{equation}
        \mathbf{\hat{R}_\uparrow} = \frac{1}{\sqrt{2}}\left(\hat{\mathbf{x}} - i\hat{\mathbf{y}}\right).
    \end{equation}
\end{subequations}
At $\mathbf{r_\downarrow}$, however, the wave travels in the $-\hat{\mathbf{z}}$ direction; an appropriately defined circular polarization basis here is
\begin{subequations}
    \begin{equation}
        \mathbf{\hat{L}_\downarrow} = \frac{1}{\sqrt{2}}\left(-\hat{\mathbf{x}} + i\hat{\mathbf{y}}\right)
    \end{equation}
    \begin{equation}
        \mathbf{\hat{R}_\downarrow} = \frac{1}{\sqrt{2}}\left(-\hat{\mathbf{x}} - i\hat{\mathbf{y}}\right).
    \end{equation}
\end{subequations}
Replacing $\hat{\mathbf{x}}\to \left(\mathbf{\hat{L}}_\uparrow + \mathbf{\hat{R}}_\uparrow\right)/\sqrt{2}$ and $\hat{\mathbf{y}}\to -i\left(\mathbf{\hat{L}}_\uparrow - \mathbf{\hat{R}}_\uparrow\right)/\sqrt{2}$ in Eq. (\ref{eq_EfieldsCart})a and rearranging yields:
\begin{eqnarray}
    \mathbf{E}(\mathbf{r_\uparrow},t) = \frac{A}{\sqrt{2}}\Bigg\{ \left( \mu_x + \frac{i m_y}{c} - i\mu_y - \frac{m_x}{c} \right)\mathbf{\hat{L}_\uparrow} \nonumber \\ + \left( \mu_x + \frac{i m_y}{c} + i\mu_y + \frac{m_x}{c} \right)\mathbf{\hat{R}_\uparrow} \Bigg\}.
\end{eqnarray}
Replacing $\hat{\mathbf{x}}\to -\left(\mathbf{\hat{L}}_\downarrow + \mathbf{\hat{R}}_\downarrow\right)/\sqrt{2}$ and $\hat{\mathbf{y}}\to -i\left(\mathbf{\hat{L}}_\downarrow - \mathbf{\hat{R}}_\downarrow\right)/\sqrt{2}$ in Eq. (\ref{eq_EfieldsCart})b yields:
\begin{eqnarray}
    \mathbf{E}(\mathbf{r_\downarrow},t) = \frac{A}{\sqrt{2}}\Bigg\{ \left( -\mu_x + \frac{i m_y}{c} - i\mu_y + \frac{m_x}{c} \right)\mathbf{\hat{L}_\downarrow} \nonumber \\ + \left( -\mu_x + \frac{i m_y}{c} + i\mu_y - \frac{m_x}{c} \right)\mathbf{\hat{R}_\downarrow} \Bigg\}.
\end{eqnarray}
The equal-time mutual coherence \cite{mandel1995optical} between the $\mathbf{\hat{L}_\uparrow}$ component of $\mathbf{E}(\mathbf{r_\uparrow},t)$ and the $\mathbf{\hat{R}_\downarrow}$ component of $\mathbf{E}(\mathbf{r_\downarrow},t)$ is:
\begin{eqnarray}
    \Gamma_{LR}(\mathbf{r_\uparrow},\mathbf{r_\downarrow}) &=& \left\langle \left[\mathbf{\hat{L}_\uparrow^\dagger}\mathbf{E}(\mathbf{r_\uparrow},t)\right]^* \left[\mathbf{\hat{R}_\downarrow^\dagger}\mathbf{E}(\mathbf{r_\downarrow},t)\right] \right\rangle \nonumber \\ &=& \frac{|A|^2}{2}\Bigg\langle -\mu_x^2 -\mu_y^2  + \frac{2i}{c}\left(\mu_x m_y - \mu_y m_x\right) \nonumber \\ &&\quad \quad \quad \quad+ \frac{m_x^2 + m_y^2}{c^2} \Bigg\rangle \nonumber \\ &=& -\frac{|A|^2}{3}\left( \mu_0^2 - \frac{m_0^2}{c^2} \right),
\end{eqnarray}
where in the last step we have carried out an average over all source orientations. The orientationally averaged intensities at these two points with the indicated polarizations are:
\begin{subequations}
    \begin{eqnarray}
        I_L\left(\mathbf{r_\uparrow}\right) &=& \left\langle \left|\mathbf{\hat{L}_\uparrow^\dagger}\mathbf{E}(\mathbf{r_\uparrow},t)\right|^2 \right\rangle \nonumber  \\ &=& \frac{|A|^2}{3} \left(\mu_0^2 + \frac{m_0^2}{c^2} - \frac{2}{c} \boldsymbol{\upmu_0}\cdot\mathbf{m_0}\right)
    \end{eqnarray}
    \begin{eqnarray}
        I_R\left(\mathbf{r_\downarrow}\right) &=& \left\langle \left|\mathbf{\hat{R}_\downarrow^\dagger}\mathbf{E}(\mathbf{r_\downarrow},t)\right|^2 \right\rangle \nonumber  \\ &=& \frac{|A|^2}{3} \left(\mu_0^2 + \frac{m_0^2}{c^2} + \frac{2}{c} \boldsymbol{\upmu_0}\cdot\mathbf{m_0}\right),
    \end{eqnarray}
\end{subequations}
and therefore the mutual coherence normalized by total intensity at those two points and polarizations is:
\begin{equation}
    \frac{\Gamma_{LR}\left(\mathbf{r_\uparrow},\mathbf{r_\downarrow}\right)}{I_L(\mathbf{r_\uparrow})+I_R(\mathbf{r_\downarrow})} = -\frac{\mu_0^2 - m_0^2/c^2}{2\left(\mu_0^2 + m_0^2/c^2\right)}.
\end{equation}
Comparing to the expressions in Eqs. (\ref{eq_generalizeddipolestrength}), (\ref{eq_chidefn}), and (\ref{eq_rhopmz}), we see that the so-normalized classical mutual coherence is exactly equal to the matrix element $\braket{1\left(k_0\mathbf{\hat{z}},L\right) |\bar{\rho}_\pm^{(\mathbf{\hat{z}}L,\hat{\mathbf{z}}R)}|1\left(k_0\mathbf{\hat{z}},R\right)}$ upon setting the quadrupole contribution to zero. Thus the corresponding coherence can be understood equally well from a classical or quantum vantage.

\bibliography{mybib}

\end{document}